\documentclass[fleqn,usenatbib]{mnras}

\usepackage{newtxtext,newtxmath}

\usepackage[T1]{fontenc}

\DeclareRobustCommand{\VAN}[3]{#2}
\let\VANthebibliography\thebibliography
\def\thebibliography{\DeclareRobustCommand{\VAN}[3]{##3}\VANthebibliography}

\usepackage{graphicx}	
\usepackage{amsmath}	

\title[Europium-enhanced stars in the Galactic disk]{Snapshots of \textit{r}-process production in the Milky Way disk}

\author[S. G. Kane et al.]{Sarah G. Kane$^{1}$\thanks{E-mail: sgk27@cam.ac.uk (SGK)}, Stephanie Monty$^{2,3}$, Madeleine McKenzie$^{4}$\thanks{NASA Hubble Fellow}, HanYuan Zhang$^{1}$, Sven Buder$^{5}$,\newauthor Terese Hansen$^{6}$, Tadafumi Matsuno$^{7}$, Jason Sanders$^{8}$, Anke Ardern-Arentsen$^{1}$, Vasily Belokurov$^{1}$,\newauthor Zofia Kaczmarek$^{7}$, and Andrew Garner$^{9}$
\\
$^{1}$Institute of Astronomy, University of Cambridge, Madingley Road, Cambridge CB3 0HA, UK\\
$^{2}$Center for Interdisciplinary Exploration and Research in Astro-
physics (CIERA), Northwestern University, 1800 Sherman Avenue,
Evanston, IL 60201, USA\\
$^{3}$Department of Astronomy, New
Mexico State University, Las Cruces, NM 88003, USA\\
$^{4}$Observatories of the Carnegie Institution for Science, 813 Santa Barbara St., Pasadena, CA 91101, USA\\
$^{5}$Research School of Astronomy and Astrophysics, Australian National University, Canberra, ACT 2611, Australia\\
$^{6}$Department of Astronomy, Stockholm University, AlbaNova University Center, SE-106 91 Stockholm, Sweden\\
$^{7}$Zentrum f{\"u}r Astronomie der Universit{\"a}t Heidelberg, Astronomisches Rechen-Institut, M{\"o}nchhofstr. 12-14, 69120 Heidelberg, Germany\\
$^{8}$Department of Physics and Astronomy, University College London, London WC1E 6BT, UK\\
$^{9}$School of Mathematics \& Physics, University of Surrey, Guildford GU2 7XH, UK
}

\date{Accepted XXX. Received YYY; in original form ZZZ}

\pubyear{\the\year{}}

\begin{document}
\label{firstpage}
\pagerange{\pageref{firstpage}--\pageref{lastpage}}
\maketitle

\begin{abstract}

Stars enhanced in heavy neutron capture elements like europium (Eu) provide a window into the rarest and most exotic nucleosynthetic events. Most Eu-enhanced stars discovered thus far are found among metal-poor populations, with $\mathrm{[Fe/H]}<-1.0$.
Here, we report the discovery of $17$ vetted Eu-enhanced stars ($\mathrm{[Eu/Fe]}>0.7$) in the Milky Way disk with metallicities in the range $-0.87<\mathrm{[Fe/H]}<-0.21$ identified from among $74\,717$ stars in the GALAH survey. 14 of these stars additionally have $\mathrm{[Ba/Eu]}<0$, marking enhancement from the rapid neutron capture ($r$-) process and making these r-II stars. Included in their number is a star with a remarkably high $\mathrm{[Eu/H]}=1.01\pm0.09$ at $\mathrm{[Fe/H]}=-0.21$. In addition to Eu, all 17 stars show enhancements in other neutron capture elements, especially in Nd, while otherwise appearing chemically consistent with typical disk stars, particularly in their $\alpha$-element abundances.
Two of the Eu-enhanced stars exhibit indications of binarity with strong enrichment in Ba and Y, marking potential signatures of the slow or intermediate neutron capture processes.
This sample of 17 Eu-enhanced stars, which span both the high- and low-$\alpha$ disks, are valuable tracers of neutron capture processes, including the $r$-process, at high metallicities. 
Although we use criteria for Eu-enhanced stars typically applied to the halo to enable comparison to literature studies of $r$-process enhanced stars, we propose a new, metallicity-dependent selection for Eu enhancement that is better-suited to the regime of the Galactic disk.
\end{abstract}

\begin{keywords}
stars: abundances -- Galaxy: disc -- Galaxy: stellar content
\end{keywords}



\section{Introduction}

The rapid neutron capture ($r$-) process is the nucleosynthetic channel responsible for creating about half of the elements in the Universe heavier than iron, including gold, silver, uranium, and thorium \citep{Arnould_2007,Cowan_2021,Thielemann_2026}. Among these elements, the abundances of which are often difficult to derive from stellar spectroscopy, europium (Eu) is most frequently used as a tracer in stars, in part because it is the $r$-process element with the strongest absorption features at optical wavelengths (especially at $\sim4129$~\AA~and~$\sim6645$~\AA). This one element has proved incredibly powerful as a window into $r$-process abundances across the stellar populations of the Milky Way \citep[MW;][]{Christlieb_2004,Barklem_2005,Hansen_2018,RPA_Sakari_2018,Gaia_eso_survey, Gaia_ESO_results,Lombardo_2025} and its nearby satellites \citep[][among many others]{ji2016a, ji2016b,  roederer16, kirby2017, venn2017, chiti2022, hansen2024, Henderson_2025}. 

Within the MW, the generally older, metal (Fe)-poor ($\mathrm{[Fe/H]}\lesssim-1$) stellar halo exhibits an approximately constant relationship of [Eu/Fe] abundances versus metallicity among its constituent stars, although these abundances have a wide variation that reflects the stochasticity of $r$-process events \citep{Haynes_Kobayashi_2019,vandeVoort2020,Cescutti_2014}. Then, at $\mathrm{[Fe/H]}\approx-1$ there is an inflection or ``knee,'' whereafter stellar [Eu/Fe] abundances begin decreasing with increasing [Fe/H] \citep{Battistini_Bensby_2016,Zhao2016}. This ``knee'' and the subsequent behavior is indicative of the onset of Fe production by Type Ia supernovae \citep[SNe,][]{Kobayashi_2020_nucleosynthesis}. The anti-correlation between [Eu/Fe] and [Fe/H] in disk stars is such that although average [Eu/Fe] values are $\sim0.5$ at $\mathrm{[Fe/H]}=-1$, they are $\sim0.25$  at $\mathrm{[Fe/H]}=-0.5$ and close to Solar at $\mathrm{[Fe/H]}=0$ \citep[see Fig. 2 in][]{Battistini_Bensby_2016}. Concurrently, the large star-to-star scatter in [Eu/Fe] seen in the metal-poor halo diminishes toward higher metallicities, with [Eu/Fe] dispersions among disk stars being quoted as $\sim0.1$ \citep{DelgadoMena_2017,AMBRE_Guiglion_2018}.
Thus, while stars with enhanced Eu abundances of $\mathrm{[Eu/Fe]}>0.7$ constitute approximately $12\%$ of the metal-poor halo \citep{Hansen_2018,RPA_Sakari_2018,Ezzeddine_2020}, such stars are exceedingly rare in the more metal-rich Galactic disk ($\mathrm{[Fe/H]}>-1$), where Eu abundances are both lower and more homogeneous.

The $r$-process occurs in extreme environments which can produce the high neutron fluxes necessary for rapid neutron capture \citep{Arnould_2007,Cowan_2021}. However, the task of retracing the history of the $r$-process in the Galaxy is complicated by the fact that the dominant site(s) of this process remain under debate. Of the proposed sites, binary neutron star mergers (NSMs) have been confirmed via spectroscopy of a kilonova \citep{Abbott_2017_NSM, Smartt_2017_rProcess,Drout_2017} immediately following the corresponding gravitational wave signal \citep{GW170817_NSM_discovery}, and magnetar flares have also been tentatively observed as an $r$-process source \citep{Patel_2025}. Although NSMs are a confirmed site of the $r$-process, many chemical evolution models nonetheless must invoke additional sites to match observed $r$-process abundances within stars. In particular, because europium abundances ([Eu/Fe]) are so elevated at low metallicities and thus early times, a more prompt enrichment source than NSMs is often demanded in explanation of observations in this regime \citep{Wehmeyer_2015,Cote_2019,Skuladottir_2020}. The most popular of these are rare classes of supernovae, including collapsars \citep{Siegel_2019,Barnes_2022} and magneto-rotational core-collapse supernovae \citep[MR-SNe;][]{Symbalisty_1984,Cameron_2003,Nishimura_2015,Reichert_2021}, although such sites for the $r$-process remain speculative. Crucially, the $r$-process in these SNe sources is thought to be highly metallicity dependent and more significant in the metal-poor regime \citep[see the discussion in][]{Fraser_2022}. Even at high metallicities the picture of the $r$-process is far from resolved, with many models continuing to require a prompt enrichment source or metallicity-dependent NSM rates to match observed Eu abundances in the Galactic disk \citep{Simonetti_2019,Chen_2025, Molero_2023, Kobayashi_2023}. Nonetheless, some works still find binary NSMs as the main $r$-process production site \citep{Fraser_2022}, for instance via more sophisticated modeling of a multi-phase interstellar medium \citep[ISM,][]{Schonrich_Weinberg_2019}.

Future gravitational wave detectors such as the Einstein Telescope \citep{Einstein_telescope_2026} will hopefully offer expanded opportunities to study binary NSMs via multimessenger astrophysics. Fortunately, stellar spectroscopic surveys and collaborations have already entered an era where the identification of $r$-process enhanced stars \textit{en masse} is possible (most recently in the $R$-Process Alliance, RPA, as seen in \citealp{Hansen_2018,RPA_Sakari_2018,Ezzeddine_2020,Holmbeck_2020,Shah_2026}; see also the reviews in \citealp{Frebel_Ji_2023, Hansen_2026}). At slightly higher metallicities in the range $-2.5\lesssim\mathrm{[Fe/H]}\lesssim -1.5$, the Measuring at Intermediate metallicity Neutron-
Capture Elements \citep[MINCE,][]{Cescutti_2022_MINCE1, Francois_2024_MINCE2, Lucertini_2025_MINCE3} project has identified several Eu-rich stars. Almost all discoveries of very $r$-process enhanced stars have thus been made among metal-poor populations with $\mathrm{[Fe/H]}<-1$, especially in the MW stellar halo or other nearby satellites. Within studies of the halo, $r$-process enhancement is split into generally three categories, defined via Eu and Ba abundances (as originally defined in \citealp{Beers_Christlieb_2005} and more recently updated in \citealp{Holmbeck_2020}). Limited-r stars are those with $\mathrm{[Eu/Fe]}<0.3$, as well as $\mathrm{[Sr/Ba]}>0.5$ and $\mathrm{[Sr/Eu]}>0.0$ \citep{Frebel_2018}. r-I stars are those with $0.3<\mathrm{[Eu/Fe]}<0.7$, and the highly enhanced r-II stars have $\mathrm{[Eu/Fe]}>0.7$. For r-I and r-II stars, an additional criterion for $\mathrm{[Ba/Eu]}<0$ is also held to ensure that stars are not Eu-enhanced due to contributions from the slow neutron capture ($s$-) process, which takes place mostly in asymptotic giant branch (AGB) stars \citep{Kappeler_2011,Lugaro_2023}. A new, rare category of extremely enhanced r-III stars have $\mathrm{[Eu/Fe]}>2$ \citep{Cain_2020, Roederer_2024}. Each of these enhanced stars bears a fingerprint of the $r$-process, enabling the study of its sites and products. Although these selection criteria are well-calibrated to metal-poor stars, they are somewhat arbitrary in the metal-rich regime. We nonetheless seek r-II stars in the disk to provide an initial benchmark study against other studies of $r$-process enhanced stars.

\begin{figure*}
    \centering\includegraphics[width=1.85\columnwidth, alt={In the left panel, the [Eu/Fe] versus [Fe/H] distribution of the kinematically selected disk stars forms a dense cloud running from the upper left (low metallicity, high Eu) to the lower right (high metallicity, lower Eu). The mean [Eu/Fe] decreases steadily with increasing metallicity, from about 0.4 at [Fe/H]=-1 to about 0.2 at [Fe/H]=-0.5, approximately 0 at Solar metallicity, and about -0.15 at [Fe/H]=0.4. The mean crosses the r-I threshold at [Eu/Fe]=0.3 near [Fe/H]=-0.7. The upper edge of the disk distribution sits at about [Eu/Fe]=0.6, with only a handful of disk stars in the grey band between 0.6 and 0.7 that marks the tolerance used to select candidates. The 17 confirmed Eu-enhanced stars sit clearly separated above this edge, with [Eu/Fe] between 0.7 and 1.22 and [Fe/H] between about -0.9 and -0.2. Most lie between [Eu/Fe] of 0.7 and 0.9, and two stand out above 1.0. The right panel shows all stars with [Fe/H] greater than -1, without the kinematic disk selection, in the space of orbital energy versus the z-component of angular momentum, with each bin colored by its mean [Eu/Fe] between 0.2 and 0.6. The stars form a V-shaped distribution, with most bins at energies between about -250000 and -120000 km^2/s^2 and angular momenta between about -1000 and 3000 kpc km/s. The bins with the highest mean [Eu/Fe], of about 0.45 to 0.6, lie above the boundary between accreted and in situ stars, at energies above about -160000 km^2/s^2 and angular momenta near 0. Below the boundary, the in situ stars have mean [Eu/Fe] of about 0.25 to 0.35. The lowest values of about 0.2 are found along the lower right edge of the distribution, where stars have the most prograde, disk-like orbits for their energy. All of the Eu-enhanced stars fall along this lower right edge, below the boundary line, at angular momenta between about 1200 and 2700 kpc km/s and energies between about -180000 and -140000 km^2/s^2. The kinematic cuts are described in Section 2.2.}]{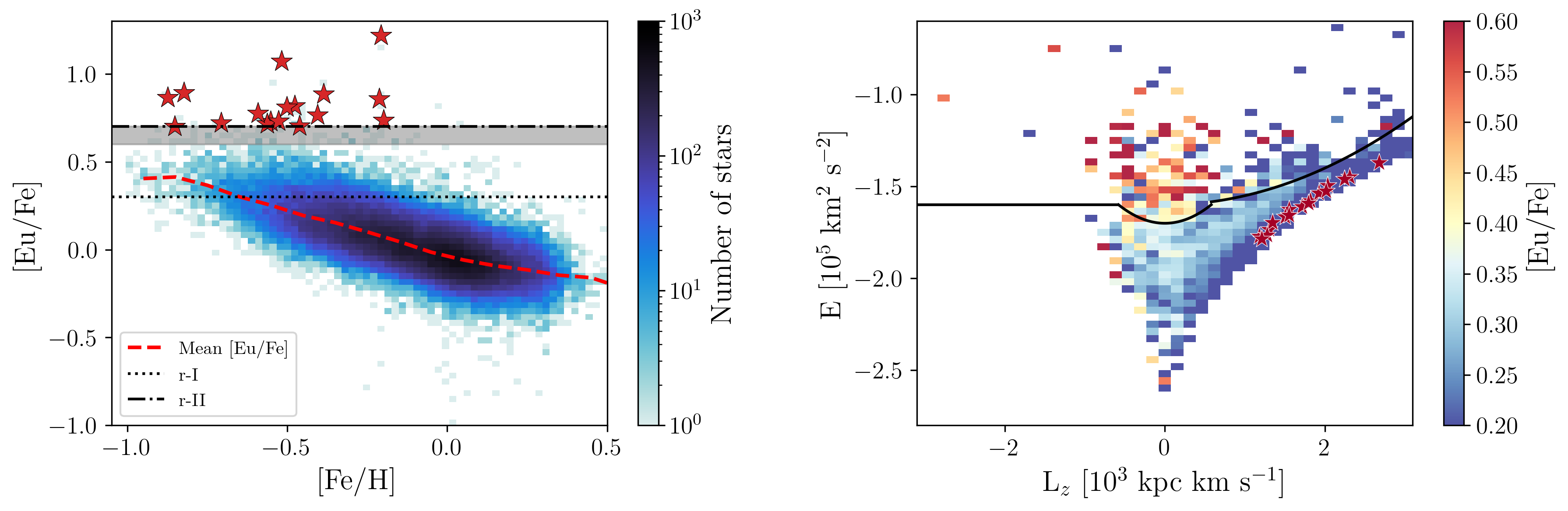}
    \caption{Left: The [Eu/Fe] versus [Fe/H] plane of all GALAH DR4 stars with $\mathrm{[Fe/H]}>-1$ and the kinematic selection of disk-like orbits (see Section~\ref{subsec:selection}). Eu abundances are adopted from GALAH DR3 or the CNN-based DR4 catalog from \citet{Kane_Kaczmarek_2026}, where applicable (see Section~\ref{subsec:data}). Selection criteria for differing levels of Eu enhancement are indicated here, with r-I enhancement marked with the dotted black line at $\mathrm{[Eu/Fe]}=0.3$ and r-II levels of Eu enhancement marked with the black dash-dotted line at $\mathrm{[Eu/Fe]}=0.7$. Our selection including stars with GALAH DR3 or CNN-based $\mathrm{[Eu/Fe]}>0.6$ is indicated with gray shading. The mean [Eu/Fe] as a function of [Fe/H] in this sample is indicated with a dashed red line. The red star markers indicate the abundances of our confirmed Eu-enhanced ($\mathrm{[Eu/Fe]}>0.7$) stars.
    Right: In the $E-L_z$ space with the MW potential from \citet{McMillan2017}, all stars passing our data quality cuts with $\mathrm{[Fe/H]}>-1$ but without kinematic selection for disk stars. Each bin is color-coded by the mean [Eu/Fe] abundance. For reference, the separation between accreted and \textit{in situ} stars from \citet{Belokurov_2023} is adapted to this potential and marked with the black solid line. Star markers are the targets in our sample confirmed as being Eu-enhanced, also color-coded by their [Eu/Fe] abundance. Note that all of the Eu-enhanced stars are on disk-like orbits in this space.}
    \label{fig:selection}
\end{figure*}

With most $r$-process enhanced stars existing in the Galactic stellar halo at low to intermediate metallicities ($\textrm{[Fe/H]}<-1$), there has been substantial work in the past decades to characterize and understand these stars. Still, metal-rich ($\textrm{[Fe/H]}>-1$) $r$-process enhanced stars remain something of a new frontier, in part due to their rarity and thus the difficulty of identifying such stars within a limited sample of Eu abundances. Nonetheless, because the $r$-process is so often invoked as metallicity dependent, having a complete picture of Eu-enhanced stars across all metallicity regimes may be essential for characterizing the sites of neutron capture events.
Pioneering work on this front has been done with \citet{Xie_2024,Xie_2025} identifying three extremely $r$-process enriched stars in the MW disk using medium resolution spectra from the Large sky Area Multi-Object fiber Spectroscopic Telescope \citep[LAMOST,][]{LAMOST, LAMOST_LEGUE}. Likewise, analysis of neutron capture abundances in disk stars from the Archéologie avec
Matisse Basée sur les aRchives de
l’ESO \citep[AMBRE,][]{Worley_2012} project also discovered five targets with $\mathrm{[Eu/Fe]}>0.7$ among stars with $\mathrm{[M/H]}>-1.0$ \citep{AMBRE_Guiglion_2018}.
As a new frontier for discovering Eu-enhancement among metal-rich populations, the GALactic Archaeology with HERMES \citep[GALAH,][]{GALAH_scientific_motivation} survey has provided Eu abundances for tens of thousands of stars via the feature at $\sim6645$~\AA, which is especially strong in metal-rich stars. Included in its most recent Data Release 4 \citep[DR4,][]{GALAH_DR4} is substantial coverage in the Galactic disk, offering a new and exciting opportunity to study the $r$-process in this environment.
Our goal is thus to provide the first systematic, large-scale study of $r$-process enhanced stars in the Milky Way disk by leveraging the vast number of stars and good disk coverage provided by GALAH DR4.

In this work, we share the discovery of $17$ new Eu-enhanced ($\mathrm{[Eu/Fe]}>0.7$) stars in the Milky Way disk at metallicities $\mathrm{[Fe/H]}>-1.0$, which are identified in the GALAH survey and confirmed via individual synthesis of the 6645\AA~Eu absorption feature. Beyond the validation of their Eu enhancement using individual re-analysis of their spectra, we also provide a preliminary chemodynamical analysis of these stars.
The paper is structured as follows: In Section~\ref{sec:data_analysis}, we outline our selection of disk stars in the GALAH survey and the identification of r-II candidates. The re-analysis of their Eu abundances via spectral synthesis is discussed in Section~\ref{subsec:synthesis}. In Sections~\ref{subsec:properties}, we discuss the other abundances of the confirmed Eu enhanced stars and place them within the context of the Galactic disk as a whole, and in Section~\ref{subsec:comparison} we compare our target stars to studies of Eu abundances in the halo and nearby dwarf galaxies.
Section~\ref{subsec:enrichment_source} discusses the potential sites of Eu production among our target stars. In Section~\ref{subsec:disk_enhanced_selection}, we propose new selection criteria for the identification of Eu-enhanced stars in the Galactic disk.
We offer a summary of our findings in Section~\ref{sec:conclusions}.

\section{Data and Analysis}
\label{sec:data_analysis}

\subsection{Eu Abundances in GALAH DR4}
\label{subsec:data}

The Eu abundances used in this work are taken as the union of the GALAH DR3 [Eu/H] abundances, where available, with the ``golden sample'' of [Eu/H] abundances from the DR4 value-added catalog developed in \citet{Kane_Kaczmarek_2026}. These datasets are adopted in preference to GALAH DR4's Eu values, since Eu abundance accuracy was reported to be lower in DR4 than in DR3.
Using [Eu/H] rather than [Eu/Fe] helps to mitigate any differences in metallicity between the two data releases, and when Fe abundances are needed to calculate [Eu/Fe], we use the GALAH DR4 [Fe/H] for all stars for consistency, although metallicities between DR3 and DR4 generally agree.\footnote{$\mathrm{[Eu/Fe]}=\mathrm{[Eu/H]}-\mathrm{[Fe/H]}$} 
Because the convolutional neural network (CNN)-derived Eu abundances from \citet{Kane_Kaczmarek_2026} are trained with [Eu/H] labels from GALAH DR3, [Eu/H] abundances from DR3 and the DR4 value-added catalog can be taken as homogeneous. Throughout this work, we apply the same selection criteria used in \citet{Kane_Kaczmarek_2026} to both the DR3 Eu abundances and the CNN-determined abundances. The goal of these criteria is to use only those stars with robust Eu abundances determined from reliable, high signal-to-noise ratio (SNR) spectra. In particular, the selection cuts applied to the DR4 catalog are:
\begin{itemize}
    \item $\texttt{flag\_sp}=0$ (for the stellar parameters)
    \item $\texttt{flag\_sp\_fit}=0$ 
    \item $\texttt{flag\_red}=0$ (for the spectrum reduction pipeline)
    \item $\texttt{snr\_px\_ccd3}>50$ (for the signal-to-noise ratio on CCD3)
    \item $T_\textrm{eff}<6600$ K
\end{itemize}
Note that a flag with a value of $0$ indicates no detected issues for that part of the analysis in GALAH. We also use only giant stars \citep[as selected via the criteria from][]{Borisov2022}, which have more reliable Eu measurements across a wider range of abundances.

To take advantage of the homogeneity between GALAH DR3's [Eu/H] values and the recomputed DR4 [Eu/H], we apply the same cuts used by \citet{Kane_Kaczmarek_2026} to develop the training and validation dataset to the data with GALAH DR3 Eu abundances when we use them:
\begin{itemize}
    \item $\texttt{flag\_fe\_h}=0$
    \item $\texttt{red\_flag}=0$
    \item $\texttt{snr\_c3\_iraf}>50$
    \item $\texttt{flag\_eu\_fe}=0$
\end{itemize}

Likewise, for consistency and to use only the most reliable CNN-determined Eu abundances, we use only the identified ``golden sample'' of the CNN abundances.
We refer the reader to Sections~4.3 and 4.6 of \citet{Kane_Kaczmarek_2026} for a detailed description of the quality cuts applied and their rationale, but in brief, the criteria for the ``golden sample'' are as follows:
\begin{itemize}
    \item $T_\mathrm{eff} > 4000 $ K, as cool stars were not well-represented in the training data and thus often had spurious [Eu/H] predictions from the CNN
    \item Normalized continuum flux $>0.93$ and $<1.07$, which excludes stars with problematic continuum normalization; this is mostly an issue for cool stars with strong molecular features
    \item $\texttt{line\_depth}>0.01$, where \texttt{line\_depth} measures the difference in normalized flux between the continuum and the Eu absorption feature at $6645.11$~\AA. This is implemented by \citet{Kane_Kaczmarek_2026} to ensure that the Eu feature is present in the spectrum beyond the underlying noise.
\end{itemize} 
We then take our final dataset as the combination of the GALAH DR3 Eu abundances and the re-computed DR4 Eu abundances passing their respective quality criteria. The final, combined set of Eu abundances includes $104\,885$ stars.

Where used, astrometry, proper motions, and radial velocities are all taken from \textit{Gaia} Data Release 3 \citep{Gaia_Overview_2016,Gaia_DR3_2023}, and we use the geometric distances from \citet{BailerJones_2021}, which are in turn derived from \textit{Gaia}. We require all stars to have distance uncertainties less than $20\%$ from \citet{BailerJones_2021}. 

All stellar parameters and other elemental abundances used are taken from the GALAH DR4 catalog. In any instance where we use an element X other than Eu, we also require that $\texttt{flag\_X\_Fe}==0$ in GALAH. Future work may re-analyze the abundances of our final sample, and non-local thermodynamic equilibrium (NLTE) corrections may be especially relevant for some abundances. We consider such re-analysis to be beyond the scope of this initial work for all elements except Eu and thus use other GALAH DR4 abundances where relevant.

\subsection{Selection of Eu-Enhanced Candidates}
\label{subsec:selection}

We apply a series of chemodynamical cuts to select both thick and thin disk stars. Consistent with the approximate metallicity of the end of spin-up (formation of the disk) in the Milky Way \citep[see][]{Belokurov_2022, Conroy_2022, Chandra_2024, Zhang_2024}, we select stars only with $\mathrm{[Fe/H]}>-1$.
We use \texttt{Astropy}'s \citep{astropy:2013, astropy:2018, astropy:2022} default Solar position and velocity within the Galactocentric reference frame, although we adopt $L_z>0$ in the direction of prograde rotation for visualization purposes, and the Milky Way potential from \citet{McMillan2017}. Orbital properties are calculated in \texttt{AGAMA} \citep{AGAMA}. Our goal is to select both thin and thick disk stars, and our kinematic cuts are designed accordingly. To identify stars on approximately circular orbits, we require $L_z/L_{z,~\textrm{circ}}(E)>0.9$ \citep{Chandra_2024}, where $L_{z,~\textrm{circ}}(E)$ is the $z$-component of the angular momentum of a star on a circular orbit at a given energy $E$. To identify stars that orbit approximately within the disk plane, we require that a star's maximum extent from the Galactic midplane during its orbit, $z_\mathrm{max}$, does not exceed $1.5~\textrm{kpc}$ \citep{BlandHawthorn_2016}. Following the metallicity cut and these kinematic selection criteria, our disk sample includes $74\,717$ stars.

The left panel of Fig~\ref{fig:selection} shows the overall [Eu/Fe] distribution of the selected disk stars as a function of [Fe/H]; note the steadily decreasing [Eu/Fe] abundance with increasing metallicity, which is also marked via the mean [Eu/Fe] abundance in the red dashed line. In the right panel of Fig~\ref{fig:selection}, there is the $E-L_z$ plane of GALAH DR4 stars with $\textrm{[Fe/H]}>-1$ color-coded by their mean [Eu/Fe] abundances; note that their is no kinematic selection for disk stars here, although disk stars will dominate at high metallicities. The stars on disk-like orbits in this space (with the highest values of $L_z$ at a given energy $E$) have consistently lower [Eu/Fe] abundances compared to non-disk stars and especially to those stars in the accreted halo. This observation is well-aligned with previous findings that the disk is typically low in Eu \citep[e.g.,][]{Battistini_Bensby_2016,Zhao2016} and that accreted stars tend to be elevated in $r$-process abundances \citep{Matsuno_2021, Aguado_2021, Monty_2024}. Note that this figure is by design very similar to the middle right panel of Fig.~14 in \citet{Kane_Kaczmarek_2026} \citep[see also Fig.~1 of][]{Monty_2024}.

\begin{figure}
    \centering\includegraphics[width=\columnwidth, alt={Both spectra include four prominent absorption features: the nickel line at about 6643.5 angstroms, the europium feature at 6645.11 angstroms, and two weaker iron lines at about 6647 and 6648 angstroms. The top panel shows the star with the lowest [Eu/H] in our sample, of about -0.15. Here the nickel line is the strongest feature, with a depth of about 0.25 in normalized flux, while the europium feature is weaker, with a depth of only about 0.08, and the iron lines are barely visible, with depths of a few hundredths. The bottom panel shows the star with the highest [Eu/H], of about 1.01. Here the nickel line is slightly shallower, at a depth of about 0.21, but the europium feature is much stronger, with a depth of about 0.17, making it nearly as deep as the nickel line. In both spectra, the synthetic fit from Korg covers a 2 angstrom window around the europium line, which includes a small portion of the continuum, and closely follows the observed flux across the entire fitted region. The uncertainties on the observed flux are small.}]{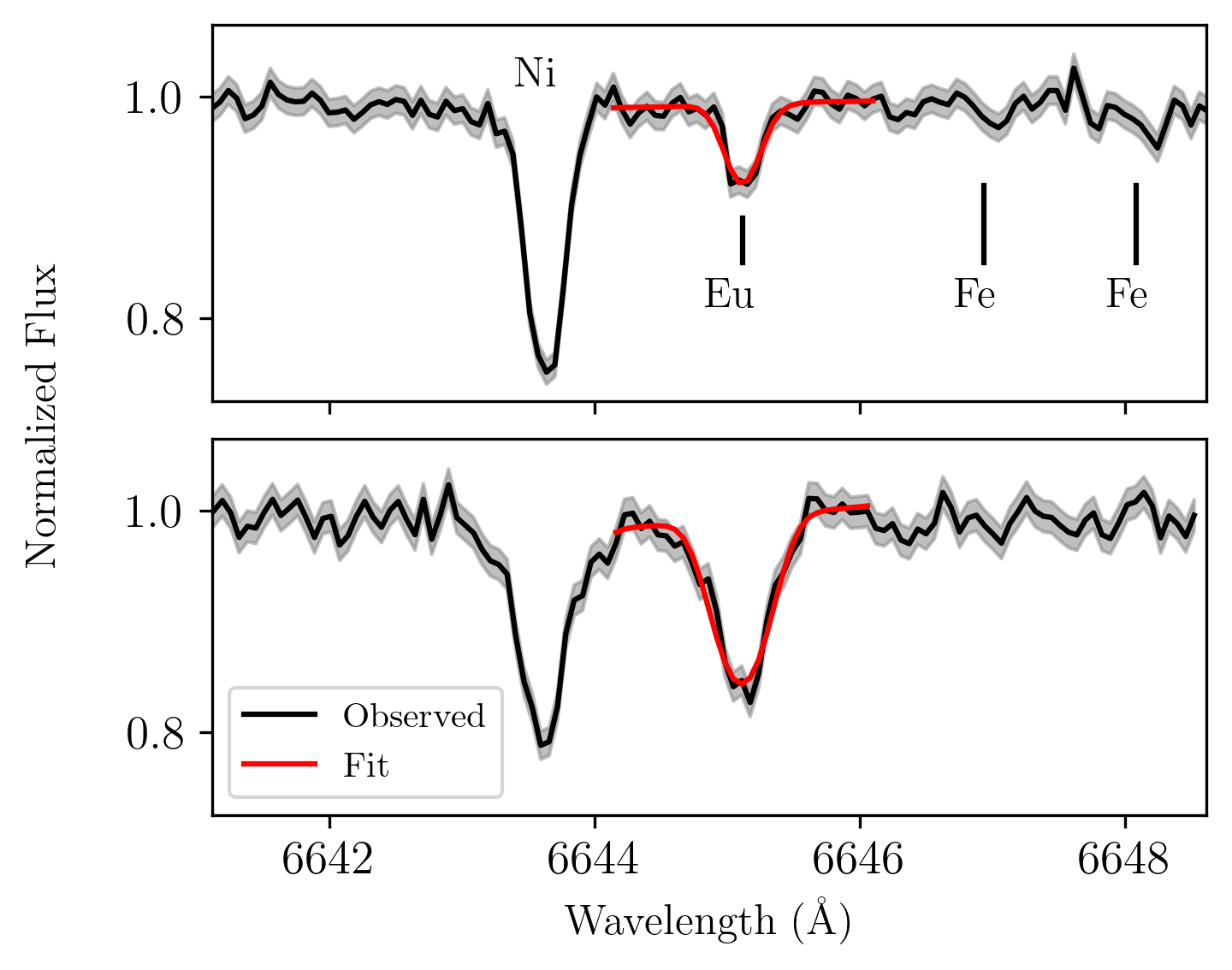}
    \caption{For illustrative purposes, two spectra from our selected r-II sample. In particular, the top and bottom panels show the spectra associated with the stars with the lowest and highest [Eu/H] abundances in our final sample, which are -0.15 and 1.01, respectively. Their corresponding [Eu/Fe] abundances are 0.70 and 1.22. The observed spectrum from GALAH DR4 is marked in black, and the gray shading indicates the uncertainties on the flux. The red line indicates the best-fit synthetic spectrum from \texttt{Korg}. The locations of the Eu feature at $6645.11$~\AA~and neighboring Ni and Fe lines, are marked.}
    \label{fig:example_spectra}
\end{figure}

Candidates are selected with the goal of identifying stars with r-II levels of Eu enrichment, defined as $\mathrm{[Eu/Fe]}\geq0.7$. However, to allow for realistic errors in the in the GALAH or CNN abundances and also because neural networks tend to ``regress towards the mean'' (e.g., in this case by underestimating high values), we allow an error tolerance of 0.1~dex, which is consistent with the typical precision reported in \citep{Kane_Kaczmarek_2026}. Thus, as illustrated by the gray shading in the left panel of Fig~\ref{fig:selection}, Eu-enhanced candidates are identified as those disk stars with $\mathrm{[Eu/Fe]}>0.6$. The right panel of Fig.~\ref{fig:selection} shows GALAH stars with $\mathrm{[Fe/H]}>-1$ in the $E-L_z$ plane, indicated as the star markers which are also colored by their [Eu/Fe] abundance. Note that the Eu-enhanced candidates are on disk-like orbits (e.g., at the most positive $L_z$ values for a given energy) and are far more enhanced in [Eu/Fe] than the surrounding disk stars. We do not yet make any requirements in [Ba/Fe] or [Ba/Eu] abundances, as we are interested in a variety of neutron capture processes enhancement (i.e., both the slow and rapid neutron capture processes). However, in Section~\ref{subsec:properties} we do examine the Ba abundances of our sample to identify those which r-II stars (with $\mathrm{[Ba/Eu]}<0$). In total, we identify 78 Eu-enhanced candidates ($\mathrm{[Eu/Fe]}>0.6$) on disk-like orbits.

\subsection{Re-Analysis of Eu Abundances}
\label{subsec:synthesis}

Our goal in re-analyzing the Eu abundances for our candidate stars is twofold: First, we wish to confirm, with as much confidence as possible from a single line, that the target stars are genuinely r-II with $\mathrm{[Eu/Fe]}>0.7$. Abundance confirmation is especially necessary when working with unusual, outlier stars, where the standard pipelines may be less reliable. Second, we wish to give an approximation as to the level of Eu-enhancement, though naturally higher resolution follow-up that includes more Eu features would be necessary to do this robustly.

To achieve these goals, we synthesize and re-fit the $6645.11$~\AA~Eu feature in the spectra with the 1D local thermal equilibrium (LTE) spectral synthesis code \texttt{Korg} \citep{KORG_2023, Korg_2024} to confirm their abundances individually. Note that this is the only reasonably strong Eu feature within the GALAH wavelength range, although there is also a much weaker absorption feature at $\sim5819$~\AA~which is nonetheless visible for $12$ stars in the final Eu-enhanced sample.
We use the GALAH DR4 spectra, which are continuum-normalized and RV-corrected with a resolution of $R=28\,000$, in a $2$~\AA~window surrounding the Eu feature at $6645.11$~\AA\footnote{In air wavelengths; note that \texttt{Korg} uses vacuum wavelengths}. We have confirmed that each of our targets has spectra from only one observation, meaning that the Eu feature could not have been artificially broadened by issues with stacking spectra. We adopt the GALAH DR4 stellar parameters ($T_\mathrm{eff}$, $\log g$, [Fe/H] for the model atmosphere metallicity, $v_\mathrm{mic}$, and $v\sin i$). To ensure that the GALAH stellar parameters are robust, we visually inspected the spectra and associated GALAH DR4 pipeline fits of the stars in our final sample\footnote{For a star with GALAH SObject ID \texttt{140809002101130}, a PDF with the observed spectrum and best fit synthetic spectrum from DR4 can be found online at \url{https://cloud.datacentral.org.au/teamdata/GALAH/public/GALAH_DR4/analysis_products_allstar/140809/140809002101130/140809002101130_allstar_fit_comparison.pdf}}. We confirm that the agreement between the observations and the fits is generally good, although one star, \textit{Gaia} DR3 \texttt{5407989761325108992}, is a confirmed binary with a non-single star (NSS) solution from \textit{Gaia} \citep{Gaia_DR3_2023} and has a moderately high $v \sin i$ value of $11.9~\rm{km/s}$. Although a high $v \sin i$ could be a signature of broadening of the lines from absorption from both components of the binary, the single-star spectrum fit from GALAH DR4 looks well-suited to the observations, and most importantly, the Eu line looks typical for a single star. For these reasons, we leave \textit{Gaia} DR3 \texttt{5407989761325108992} in our sample with these notes for caveats.

For completeness, we also checked the $T_\mathrm{eff}$ and $\log g$ of the stars in our final sample with photometry and explore the influence of these re-computed stellar parameters on the derived [Eu/H] abundances. This is described in detail in Appendix~\ref{appendix:stellar_params}.
For the synthesis with \texttt{Korg}, we use MARCS model atmospheres \citep{Gustafsson_2008} which are plane-parallel for stars with $\log g>3.5$ and spherical for those with $\log g<3.5$, which constitute the majority of our sample (which all have $\log g<3.8$). $\alpha$-element abundances in the model atmospheres are set to be consistent with GALAH DR4 \citep{GALAH_DR4}, with $\mathrm{[\alpha/Fe]}=0.4$ at $\mathrm{[Fe/H]}=-1.0$, $\mathrm{[\alpha/Fe]}=0.0$ for $\mathrm{[Fe/H]}\geq0$, and $\mathrm{[\alpha/Fe]}$ interpolated linearly between $0.4$ and $0$ for $-1.0<\mathrm{[Fe/H]}<0.0$. Our line list is generated from \texttt{linemake} \citep{Linemake}\footnote{See the GitHub at \url{https://github.com/vmplacco/linemake}.}, with the relevant atomic data for Eu II originating in \citet{Lawler_2001}. Eu isotopic ratios are adopted from National Institute for Standards and Technology (NIST)\footnote{\href{https://www.nist.gov/pml/atomic-weights-and-isotopic-compositions-relative-atomic-masses}{NIST}}, in accordance with the \texttt{Korg} defaults \citep{KORG_2023}. The NIST isotopic ratios are $47.81\%~^{151}\rm{Eu}$ and $52.19\%~^{153}\rm{Eu}$, which reflects the abundances most typically seen in laboratory materials. We found that the \texttt{linemake} line list was able to produce a better match to the observed spectra than when the synthesis was performed with the GALAH line list \citep{GALAH_DR3}, which is modified from the Gaia-ESO line list \citep{Heiter_2015,Heiter_2020}. Although the GALAH spectra are continuum-normalized, we allow \texttt{Korg} to adjust the continuum level of its synthetic spectra when re-determining the Eu abundance, as we found this to produce more realistic fits to the data. 

Uncertainties on the [Eu/H] abundances are derived by individually adjusting the stellar parameters ($T_\mathrm{eff}$, $\log g$, [Fe/H], $v_\mathrm{mic}$) to the upper and lower bounds indicated by their GALAH DR4 uncertainties and re-determining the [Eu/H] abundance via the same method described above. The uncertainty in [Eu/H] is then taken as the sum in quadrature of the change in the abundance produced by each of the stellar parameter uncertainties. The uncertainties in [Eu/H] for the Eu-enhanced stars range between 0.09 and 0.13 (see Table~\ref{tab:eu_enhanced_stars}). The uncertainty in [Eu/Fe] is then taken as the sum in quadrature between the [Eu/H] uncertainty and the [Fe/H] uncertainty reported by GALAH.

Using the synthesis from \texttt{Korg}, we estimate the Eu abundance [Eu/H] and calculate the [Eu/Fe] values using GALAH DR4 metallicities. [Eu/H] abundances are reported on the \citet{Grevesse_2007} Solar scale for consistency with GALAH. Any candidates falling below the r-II selection threshold at $\textrm{[Eu/Fe]}=0.7$ are rejected from our final sample. From the $78$ candidates originally identified in Section~\ref{subsec:selection}, $19$ pass this first step; an additional $46$ of the stars rejected had $0.6<\mathrm{[Eu/Fe]}<0.7$ from the CNN or GALAH DR3, in accordance with our 0.1~dex error tolerance. Nonetheless, the substantial number of false positives here underscores the importance of independently validating the performance of any bulk pipeline on outlier cases. We then visually inspect the fit of the remaining synthetic spectra to ensure consistency with the observed GALAH spectrum and reject any obviously unsuitable fits; two additional stars are rejected following this visual inspection. 
Examples of our synthetic spectra fits with \texttt{Korg} alongside the corresponding observed spectra are shown in Fig.~\ref{fig:example_spectra}. After re-analysis, 17 Eu-enhanced stars with $\mathrm{[Eu/Fe]}>0.7$ remain in our final sample.

\section{Results}
\label{subsec:results}

\begin{figure}
    \centering\includegraphics[width=0.9\columnwidth, alt={The figure shows three histograms of [Fe/H] on a logarithmic scale of the number of stars, using bins of 0.1 dex. The full disk sample spans [Fe/H] from -1 to about 0.7 and peaks at roughly 10,000 stars per 0.1 dex bin between [Fe/H] of about -0.4 and 0.1. The number of stars falls steeply toward low metallicity, to about 40 stars in the most metal-poor bin between -1.0 and -0.9, and even more steeply toward high metallicity, with only a handful of stars above [Fe/H]=0.5. By contrast, the 78 r-II candidates are concentrated at lower metallicities, with about 3 to 17 candidates per bin between [Fe/H]=-1.0 and -0.1, and the most candidates between -0.6 and -0.5. Only two candidates have [Fe/H] greater than -0.1, and only one is at super-Solar metallicity. The 17 confirmed Eu-enhanced stars span a narrower range still, from [Fe/H]=-0.87 to -0.20, with 1 to 6 stars per bin and the largest group, of about 6 stars, between -0.6 and -0.5. Thus, although thousands of disk stars have super-Solar metallicities, none of the confirmed Eu-enhanced stars are found in this regime.}]{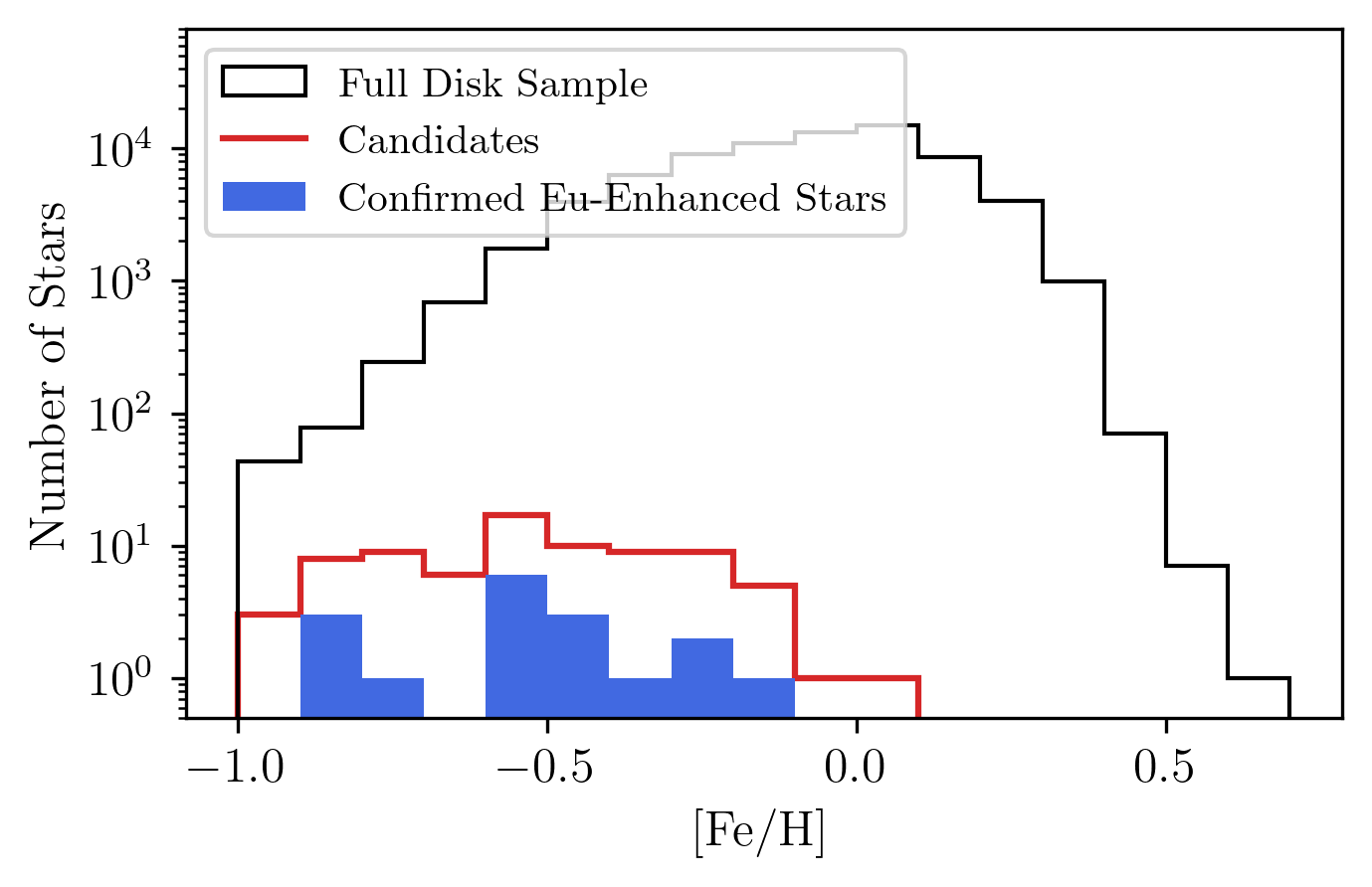}
    \caption{Metallicity ([Fe/H]) distributions of the full disk sample in the black outlined histogram, r-II candidates in the red dash outlined histogram, and confirmed r-II stars in the blue shaded histogram.}
    \label{fig:feh_dist}
\end{figure}

\begin{figure*}
    \centering\includegraphics[width=2\columnwidth, alt={In each panel, the typical high alpha disk stars with [X/Fe] about 0.2 are more dominant at lower metallicities closer to -1.0 and -0.5, while the low alpha disk sequence with [X/Fe] closer to 0 spans up to higher metallicities (to Solar and beyond). In Mg, Si, and Ti, the alpha-element bimodality is evident, although it is most prominent in Ti; in Ca, the bimodality is not clearly present. Across all four alpha elements (Mg, Si, Ca, and Ti), the Eu-enhanced stars appear broadly consistent with the overall distribution of disk stars. The Eu-enhanced stars all have [Fe/H] between about -0.9 and -0.2 and [X/Fe] between about -0.1 and 0.35, and the most metal-poor targets have the highest alpha-element abundances, of about 0.25 to 0.35, consistent with the presence of the high alpha disk at low metallicities. Most targets sit within the disk distribution or along its lower edge, and in each element two to four stars fall on the low tail. However, these stars are not consistently low across all four elements, and there is no clear trend between the alpha-element abundances of the targets and how strongly they are enhanced in Eu. The typical uncertainties are less than or equal to 0.1 in [X/Fe] and [Fe/H]. In the [Ti/Fe] panel, the line separating the high- and low-alpha disks sits at [Ti/Fe]=0.2 for [Fe/H] less than -0.65, decreases linearly to 0.1 at [Fe/H]=-0.15, and stays at 0.1 at higher metallicities (Equation 1). The Eu-enhanced stars fall on both sides of this line: seven of the r-II stars are in the high-alpha disk and seven are in the low-alpha disk, while the three other Eu-enhanced stars all fall in the low-alpha disk.}]{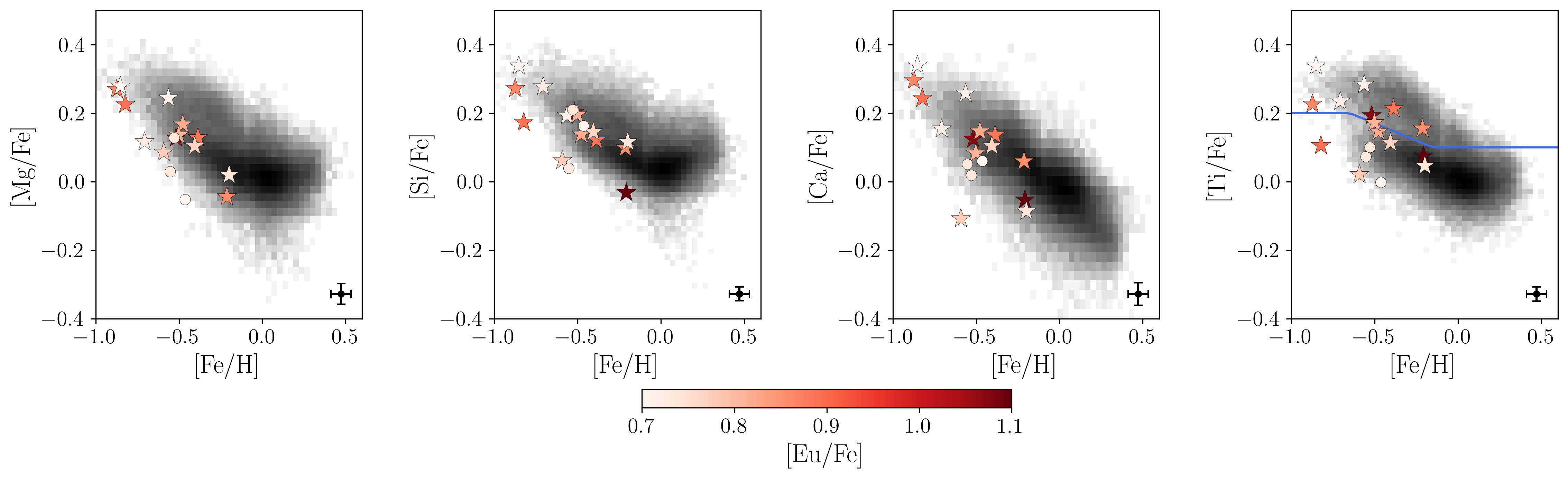}
    \caption{In the gray 2D histograms, the distribution of kinematically-selected disk stars in GALAH DR4 in the [X/Fe] versus [Fe/H] space, where X is an $\alpha$ element. From left to right, X is Mg, Si, Ca, and Ti. Abundances are adopted as the GALAH DR4 values, using only those stars where the error flag for the element X is not raised. Overplotted as star markers are the r-II stars, and the remaining Eu-enhanced stars are marked as circles, color-coded by their [Eu/Fe] abundance as determined from the synthesized [Eu/H] and the GALAH DR4 [Fe/H]. The horizontal and vertical errorbars in the lower right corner of each panel indicate the mean error reported by GALAH in [Fe/H] and [X/Fe], respectively, among the Eu-enhanced stars. In the right panel, which shows [Ti/Fe] versus [Fe/H], we have indicated an approximate separation between the high- and low$\alpha$ disks with a blue line.}
    \label{fig:alpha_elements}
\end{figure*}

\begin{figure*}
    \centering\includegraphics[width=2\columnwidth, alt={Compared to the normal disk stars, the Eu-enhanced stars appear typical or mildly enhanced in Ba and Y, but clearly enhanced in Nd. In the [Ba/Fe] panel, the disk stars mostly fall between about -0.3 and 0.5, and most of the r-II stars have [Ba/Fe] between 0 and 0.5, in the upper part of the disk distribution, with the highest at about 0.75. In the [Y/Fe] panel, most of the r-II stars have approximately Solar [Y/Fe] between about -0.2 and 0.2, similar to disk stars at the same metallicity, with one exception at about 0.75. In the [Nd/Fe] panel, the disk stars are mostly between about -0.2 and 0.4, while most of the r-II stars are between about 0.1 and 0.7 and reach as high as about 1.05. The r-II stars have low [Ba/Eu] of between about -0.85 and -0.2. The three other Eu-enhanced stars, which have [Ba/Eu] of about 0.2 to 0.7, stand out at high values in all three elements. Two of these, which show signs of binarity, have [Ba/Fe] of about 1.35 to 1.4, [Y/Fe] of about 1.0 to 1.1, and [Nd/Fe] of about 1.3. The third has [Ba/Fe] of about 0.9, [Y/Fe] of about 0.65, and [Nd/Fe] of about 1.05. In the [Y/Ba] panel, the disk stars range from about -0.5 to 0.25, and almost all of the Eu-enhanced stars have typical [Y/Ba] between about -0.5 and 0.05. The two binary stars have [Y/Ba] of about -0.25 and -0.4, and one r-II star has an unusually low [Y/Ba] of about -0.7. The typical uncertainties are less than 0.1 in [Fe/H] and [X/Fe].}]{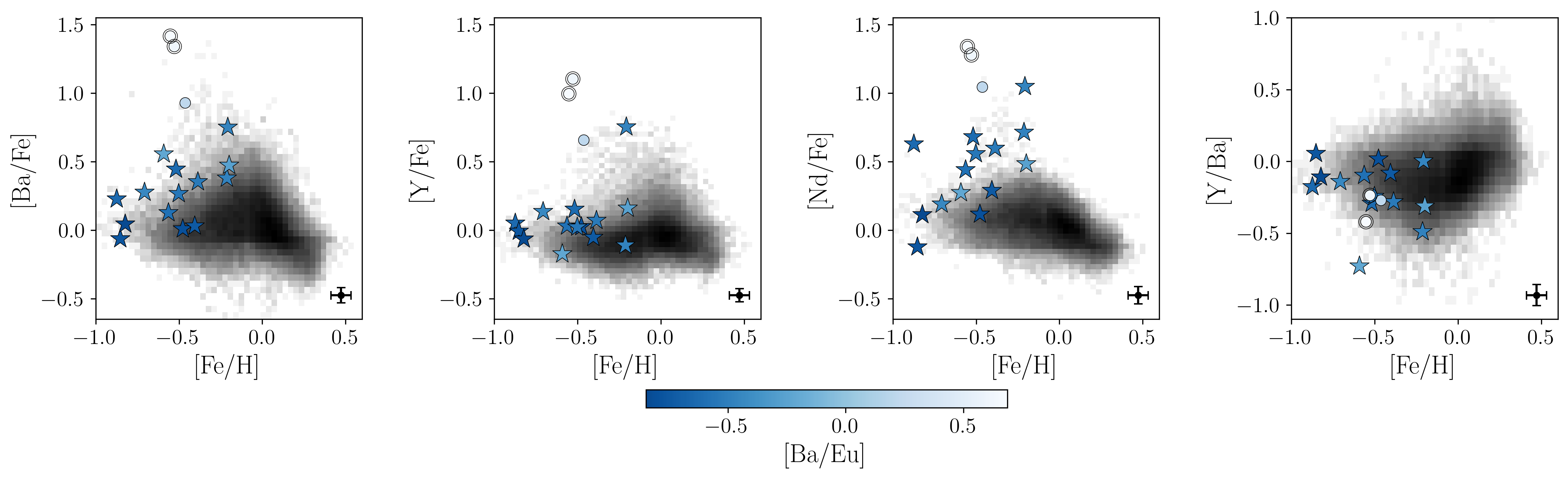}
    \caption{As in Fig.~\ref{fig:alpha_elements}, we show here the GALAH DR4 [X/Fe] versus [Fe/H] plane, now for the neutron capture elements, Ba, Y, and Nd, organized in order of increasing $r$- to $s$-process contribution at Solar. The final panel on the right gives the light-to-heavy $s$-process ratio, [Y/Ba], versus [Fe/H] abundances. The gray 2D histograms again show the overall distribution of GALAH DR4 disk stars in this space, and the overplotted star and circle markers indicate the abundances of the r-II and other Eu-enhanced stars. The two stars which are most [Ba/Fe] enhanced and exhibit signatures of binarity are also indicated via the presence of a second circle. Here, the Eu-enhanced candidates are color-coded by their [Ba/Eu] abundance, using the Eu value determined with \texttt{Korg} and GALAH DR4 Ba values. The errorbars in the lower right corner of each panel give the average error reported by GALAH for the Eu-enhanced stars.}
    \label{fig:neutron_capture}
\end{figure*}

\subsection{Chemical Properties of Eu-Enhanced Stars}
\label{subsec:properties}

Of the 17 vetted Eu-enhanced disk stars identified in the GALAH sample, 14 also have $\mathrm{[Ba/Eu]}<0$, making them confirmed r-II stars. For the remainder of this work, we distinguish the 14 true r-II stars from those that are merely Eu-enhanced (i.e., having $\mathrm{[Eu/Fe]}>0.7$ but $\mathrm{[Ba/Eu]}>0$). Accordingly, in subsequent figures these two groups are marked separately as star markers for r-II stars and circle markers for the other three Eu-enhanced stars.
Our 14 confirmed r-II stars occur at a rate of just $0.019\%$ stars within our disk sample having $\mathrm{[Eu/Fe]}>0.7$ as compared to $\sim12\%$ among metal-poor ($\mathrm{[Fe/H]}<-1$) halo stars \citep[as noted in RPA studies; see][]{Hansen_2018,RPA_Sakari_2018,Ezzeddine_2020}. By contrast, the occurrence rate of targets with $\mathrm{[Eu/Fe]}>0.7$ among stars with $\mathrm{[M/H]}>-1.0$ analyzed by the AMBRE project \citep{AMBRE_Guiglion_2018} is over an order of magnitude higher, at $0.35\%$.\footnote{Note that this rate comes from our own calculation using their public catalogs.} The fact that this occurrence differs so greatly from ours using the GALAH sample suggests either that a.) our samples are somehow very different from each other, b.) that there is a bias in our analysis of the Eu-enhanced stars, or c.) that we have missed some genuinely Eu-rich stars in our selection. Although any of these is possible, we think the last is especially likely given that our initial selection is made with a neural network, which tend to be naturally biased away from outlier values. These caveats mean that the occurrence rates quoted here for Eu-enhanced stars in the disk are likely at best lower limits, and there may be more metal-rich r-II stars in GALAH DR4 that are lost by our selection criteria.

As shown in the left panel of Fig.~\ref{fig:selection}, many of our r-II stars have [Eu/Fe] close to the classification boundary at $0.7$, but $8$ have $\mathrm{[Eu/Fe]}>0.8$ and $2$ have $\mathrm{[Eu/Fe]}>1.0$ \citep[which has in the past been used as the r-II classification cutoff;][]{Beers_Christlieb_2005}. The rate of stars with $\mathrm{[Eu/Fe]}>1.0$ is thus a mere $0.0027\%$ within our disk sample, over $1\,000$ lower than the $5\%$ rate for such stars in the halo \citep{Barklem_2005,Hansen_2018,Frebel_Ji_2023}. The most Eu-enhanced star in our sample has a remarkable $\mathrm{[Eu/Fe]}=1.22\pm0.11$ at $\mathrm{[Fe/H]}=-0.218$; to our knowledge, this is the star with the highest [Eu/H] yet known at $1.01$. The Eu feature in the GALAH spectrum for this star is shown in the bottom panel of Figure~\ref{fig:example_spectra}.

The rate of Eu-enhanced stars in the disk clearly has a strong metallicity dependence and decreases with increasing [Fe/H]. In Figure~\ref{fig:feh_dist}, we show the metallicity distribution of the 17 enhanced stars. These stars span a range of metallicities from $-0.87\leq\mathrm{[Fe/H]}\leq-0.20$; the confirmed r-II subset spans this same metallicity range. Although there are thousands of stars in GALAH with $\mathrm{[Fe/H]}\geq0$, there are no Eu-enhanced stars with super-Solar metallicity and only one candidate, which is ultimately rejected with synthesis. Likely the lack of enhanced stars at $\mathrm{[Fe/H]}\geq0$ arises because the mean [Eu/Fe] abundance decreases steadily with increasing metallicity due to the dominance of Fe production via Type Ia SNe \citep{Kobayashi_2020_nucleosynthesis}. By $\mathrm{[Fe/H]}\approx0$, the mean [Eu/Fe] in our GALAH disk sample is $\sim-0.037$, and a star with $\mathrm{[Eu/Fe]}=0.7$ would be a $7.4\sigma$ outlier. As is evident in Fig.~\ref{fig:feh_dist}, there are outliers from the disk Eu distribution in the super-Solar metallicity regime, but they have $\mathrm{[Eu/Fe]}\approx0.5$ and thus do not meet the criteria from r-II levels of Eu enhancement. Because the selection for r-II stars is essentially arbitrary in this regime, we give further attention to the need for a new, metallicity dependent selection of $r$-process enhanced stars at high metallicities in Section~\ref{subsec:disk_enhanced_selection}.

\subsubsection{$\alpha$-Element Abundances}

In Fig.~\ref{fig:alpha_elements}, we show the $\alpha$-element abundances of our Eu-enhanced stars in relation to the abundances of the overall disk sample in GALAH DR4. Several observations are apparent from the figure. First, our target stars appear broadly chemically consistent with other disk stars in Mg, Si, Ca, and Ti, suggesting a true \textit{in situ} disk origin. In each $\alpha$-element space, there do appear to be between $2$ to $4$ stars that lie on the lower tail of the abundance distribution for other disk stars. However, these stars are not consistently depleted across all $\alpha$-element spaces (e.g., those stars which are relatively low in [Mg/Fe] appear typical in [Si/Fe] and [Ti/Fe]), so we suggest that rather than being genuinely $\alpha$-depleted in relation to other disk stars, these few stars are merely slightly anomalous in individual elements. Second, there is no clear correlation between $\alpha$-element abundance and the level of enhancement in [Eu/Fe]. We discuss this in more detail below in Sec.~\ref{subsec:enrichment_source}, but there is no obvious signature of $\alpha$ enhancement coupled to the enrichment in Eu.

Finally, the Eu-enhanced stars span both the high- and low-$\alpha$ disks, although in differing numbers. An approximate separation of the $\alpha$-element bimodality in the Galactic disk is made using [Ti/Fe], as is marked in the left panel of Fig~\ref{fig:alpha_elements}. This separation, which was made by visual inspection of the bimodality in the [Ti/Fe]-[Fe/H] plane, is as follows:
\begin{equation}
[\mathrm{Ti}/\mathrm{Fe}] =
\begin{cases}
0.2 & \text{for } [\mathrm{Fe}/\mathrm{H}] < -0.65, \\
0.1 & \text{for } [\mathrm{Fe}/\mathrm{H}] \ge -0.15, \\
-0.2\times[\mathrm{Fe}/\mathrm{H}] + 0.07 & \text{else}
\end{cases}
\end{equation}

We elect to use Ti as a separator between the high- and low-$\alpha$ disks because it shows the strongest bimodality in GALAH DR4, while other $\alpha$-elements (e.g., Ca, see the third panel of Fig.~\ref{fig:alpha_elements}) do not show such a clear bimodality. Nonetheless, we recognize that this separation is approximate and only use it here to indicate that the Eu-enhanced stars span both disks in terms of chemistry. Using this separation (applied only to those stars with $\texttt{flag\_ti\_fe}=0$ in the GALAH catalog), seven r-II stars belong to the high-$\alpha$ disk and seven to the low-$\alpha$ disk. These numbers correspond to an r-II occurrence rate of $0.062\%$ and $0.011\%$ in the high- and low-$\alpha$ disks, respectively. All three of the other Eu-enhanced stars belong to the low-$\alpha$ disk.

\subsubsection{Neutron-Capture Abundances}

In Figure~\ref{fig:neutron_capture}, we show neutron-capture abundances of the Eu-enhanced stars in relation to the overall distribution of typical disk stars. The elements shown, Ba, Y, and Nd, are organized in the first three panels of the figure in order of decreasing $s$- to $r$-process contributions at Solar metallicity according to \citet{Prantzos_2020}. The fraction of the Solar abundance of each of these elements contributed by the $s$-process is 0.888 for Ba, 0.778 for Y, and 0.615 for Nd \citep[see Table 4 of][]{Prantzos_2020}. 
For comparison, the fraction of  $s$-process contribution for Eu to the Solar abundance is only $0.049$, as compared to $0.951$ from the $r$-process. Each of the enhanced stars is color-coded by their [Ba/Eu] abundance, to give an approximation of $s$-/$r$-process enrichment. Compared to the distribution of typical disk stars, which have a median [Ba/Eu] of $0.04$, most of the Eu-enhanced candidates exhibit a lower [Ba/Eu] with a median value of $-0.53$, suggesting a strong \textit{r}-process abundance and a lower but non-negligible \textit{s}-process contribution. For comparison, the pure Solar $r$-process [Ba/Eu] value is $-0.9$ following from \citet{Prantzos_2020} and $-0.8$ from \citet{Bisterzo_2014} \citep[see also Section 4.1 of][]{Sitnova_2025}. 

With regards to the two dominant $s$-process tracers, Ba and Y, many of the Eu-enhanced stars, including the r-II stars with $\mathrm{[Ba/Eu]}<0$, appear mildly enhanced compared to the disk average abundances. Nonetheless, most of the Eu-enhanced stars still lie within the typical Ba and Y abundance regime of disk stars, with all but four of the target stars approximately Solar in [Y/Fe]. In [Ba/Fe] most values for the Eu-enhanced disk stars range between $0$ and $0.5$. We suggest that the r-II stars may be slightly enhanced in Ba and Y compared to other disk stars because these elements, although produced predominantly in the $s$-process, do have some $r$-process contribution at Solar metallicity \citep[e.g., Ba is $\sim11.2\%$ $r$-process at Solar, per][]{Prantzos_2020}. Thus, extreme $r$-process enrichment may be sufficient to slightly enhance Ba and Y relative to non-enriched disk stars, but we discuss this in more detail in Section~\ref{subsec:enrichment_source}.

\begin{figure*}
    \centering\includegraphics[width=2\columnwidth, alt={The Eu-enhanced disk stars, with [Fe/H] between about -0.9 and -0.2, are much more metal-rich than essentially all of the r-Process Alliance (RPA) halo stars and many of the dwarf spheroidal galaxy (dSph) stars. The RPA stars mostly have [Fe/H] between about -3.3 and -2, while the dSph stars span from about -3.3 up to about almost 0. In the [Mg/Fe] panel, the RPA stars have [Mg/Fe] of about -0.2 to 1.5, with most between 0.3 and 1.0. The dSph stars show [Mg/Fe] decreasing from about 0.5 at [Fe/H]=-3 to about -0.2 near [Fe/H]=-0.5, so they reach the Solar value at a much lower metallicity than the disk stars. The Eu-enhanced disk stars, at [Mg/Fe] of about 0 to 0.3, are clearly more Mg-rich than dSph stars at the same metallicity. In the [Ba/Fe] panel, the RPA and dSph stars range from about -1.4 to 1.5, mostly between -0.6 and 0.5. Most of the Eu-enhanced disk stars have [Ba/Fe] of about 0 to 0.8, near the upper end of the dSph distribution at similar metallicity. The two Ba-rich binary stars reach about 1.4, similar to several of the most Ba-rich RPA stars. In the [Eu/Fe] panel, all of the disk stars sit above the r-II threshold at [Eu/Fe]=0.7, between about 0.7 and 1.2. Many of the RPA and dSph stars fall between about 0 and 0.7, but a number of RPA stars also lie above the threshold, reaching as high as about 2.2 near [Fe/H]=-3.2. Only a few dSph stars with [Fe/H] greater than -1 exceed it. In the [Eu/Ba] panel, the RPA and dSph stars mostly have [Eu/Ba] between about 0 and 1, and the r-II disk stars have [Eu/Ba] of about 0.2 to 0.85, in the upper part of this range. The three Eu-enhanced disk stars that are not r-II have negative [Eu/Ba] of about -0.2 to -0.7, and several RPA stars show similarly low values. Downward arrows mark upper limits for some of the RPA and dSph stars, mostly in [Eu/Fe] and [Eu/Ba].}]{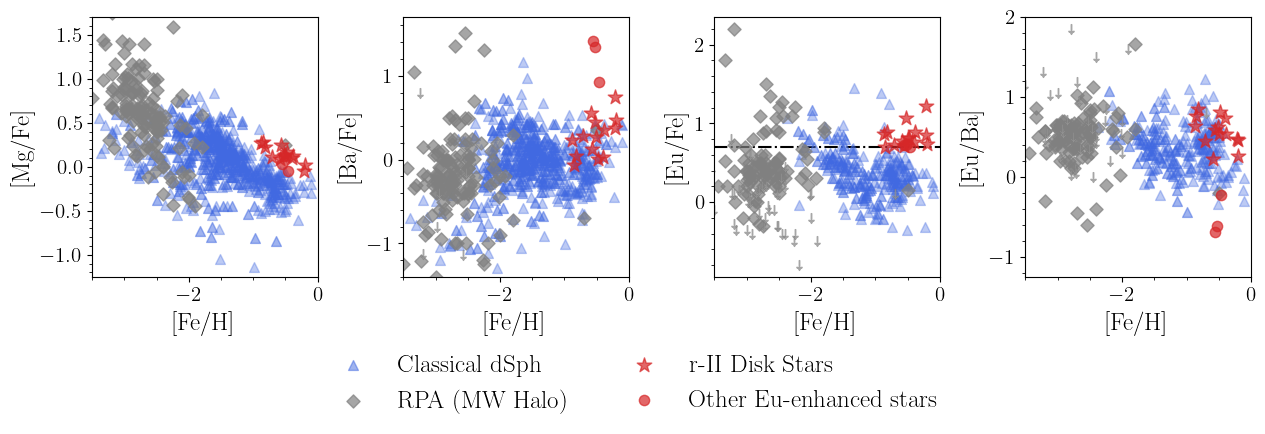}
    \caption{From left to right, the [Mg/Fe], [Ba/Fe], [Eu/Fe], and [Eu/Ba] versus [Fe/H] abundances for MW halo stars observed by the $r$-Process Alliance (RPA) and dwarf spheroidal (dSph) galaxy member stars in gray diamond and blue triangle markers, respectively. The RPA data is collected from \citet{Ezzeddine_2020}. The dwarf galaxy abundances are collected from a range of literature sources and include dSph galaxies Bootes~I, Carina, Draco, Fornax, Leo~I, Sculptor, Sextans, and Sagittarius \citep{norris2017b, hill2019, theler2020, reichert2020}. Where applicable, downward arrow markers indicate upper limits on abundances. As red star or circle markers, we also indicate the chemical properties of our Eu-enhanced disk stars using the GALAH DR4 abundances and the [Eu/H] values from synthesis. In the [Eu/Fe] versus [Fe/H] panel, we also indicate the r-II selection at $\mathrm{[Eu/Fe]}>0.7$ as the black dash-dotted line (see also Fig.~\ref{fig:selection}).}
    \label{fig:abundance_comparison}
\end{figure*}

In the first two panels of Fig.~\ref{fig:neutron_capture} showing [Ba/Fe] and [Y/Fe] abundances, two of the Eu-enhanced stars with especially elevated $s$-process abundances (both Ba and Y) are apparent. These two stars have \textit{Gaia} DR3 IDs 
\texttt{5707969656161149440} and 
\texttt{5251879932203856384} and are indicated by two concentric circle markers in Fig.~\ref{fig:neutron_capture}. Correspondingly, their $s$- to $r$-process contributions as traced by [Ba/Eu] are significantly higher than those for our other target stars at $0.69$ and $0.61$. One of these stars is also identified as Ba-rich in GALAH by \citet{Levine_2026}. In addition to the neutron capture elements illustrated here, these two stars are elevated in Ce in GALAH DR4, with [Ce/Fe] abundances of $1.52$ and $1.31$. Examination of the astrometry and radial velocities of these two stars shows that they have high re-normalized unit weight error (RUWE) values of $4.97$ and $1.48$, consistent with these stars belonging to unresolved binaries \citep{Belokurov_2020}. As further indications of binarity, \textit{Gaia} \texttt{5251879932203856384} has a \textit{Gaia} RV error of $35.5~\mathrm{km/s}$, indicative of substantial RV variation, and
\texttt{5707969656161149440} has a non-single star solution in DR3 \citep{Holl_2023,Halbwachs_2023}. Accordingly, it is probable that these two stars were $s$-process enriched by a mass transfer from a companion during its AGB stage; this companion has since progressed to become a white dwarf. Further corroborating this hypothesis is that both \textit{Gaia} DR3 
\texttt{5707969656161149440} and 
\texttt{5251879932203856384} are moderately enhanced in C, as would be expected of AGB mass-transfer products, with [C/Fe] values of $0.41$ and $0.35$ in GALAH DR4, respectively. These abundances mark them as the most C-rich of our Eu-enhanced sample, and the median [C/Fe] abundance of disk stars with similar metallicities is $0.08$.

We propose that there are two potential scenarios for the dual Eu- and Ba-enrichment of these two stars. One possibility is that these stars formed from initially $r$-process rich material alongside their binary companion, which later separately enriched them in $s$-process elements during the AGB phase. The other possibility is that the Ba and Eu enrichment occurs in the same site and thus are coupled in either the $s$-process or the  ``intermediate'' ($i$-) neutron capture process \citep{Cowan_Rose_1977}.
Heavy element abundance patterns seen in some carbon-enhanced metal-poor \citep[CEMP;][]{Beers_1992, Masseron_2010, Aoki_2007} stars support this scenario, with so-called CEMP-r/s stars enriched in both Ba and Eu in addition to the nominal C enhancement \citep[e.g.,][]{Hill_2000,Cohen_2003,Beers_Christlieb_2005}. In these CEMP-r/s stars, the abundance patterns are consistent with enrichment from an AGB companion, including the Eu enhancement, and the $i$-process is often invoked to explain the full abundance pattern \citep{Hampel_2016, Karinkuzhi_2021, Riyas_2026}.

Although the $i$-process is often studied in metal-poor AGB stars \citep{Choplin_2021, Karinkuzhi_2021}, \citet{Karinkuzhi_2023} also suggest it could occur at near-Solar metallicities. Interestingly, two of the stars in their work which they attribute to $i$-process enrichment have $\textrm{[Eu/Fe]}>0.7$ as well as enhanced Ba, very similar to our two stars here; it is thus plausible that Gaia DR3 \texttt{5707969656161149440} and 
\texttt{5251879932203856384} are two further examples of a metal-rich, $i$-process enhanced class resulting from pollution via their AGB companions.
Further, detailed spectroscopy to determine more heavy element abundances and isotopic ratios for these especially unusual stars may resolve this picture between these two scenarios, as the $i$-process may not be able to produce heavy actinides like Th and U at high metallicities \citep{Choplin_2025}, and thus these elements may only be enhanced compared to other disk stars if they are truly $r$-process enriched. Ba-isotope ratios have also been used in CEMP-r/s stars to identify the signatures of the $i$-process \citep{Sitnova_2026}, which may be possible with follow-up spectroscopy.

Two other stars also appear somewhat Y-rich in comparison to the rest of the disk. One of these stars has $\mathrm{[Ba/Eu]}=0.22$, making it another of the Eu-enhanced but not r-II stars, while the other Y-rich star has a much lower $\mathrm{[Ba/Eu]}=-0.47$, which is typical of the other r-II stars. This strangely Y-enhanced but Ba-typical star is also evident in the [Y/Ba] plane in the right panel of Fig.~\ref{fig:neutron_capture}. Neither of these stars has elevated \texttt{RUWE} values, although this cannot entirely rule out the possibility of binarity. Two other stars in our sample, \textit{Gaia} \texttt{5407989761325108992} and \texttt{5815617720050615296}, are also flagged as non-single stars in \textit{Gaia} DR3 but appear normal among our Eu-enhanced sample among their $s$-process tracer elements, particularly Ba and Y. We thus suggest that for these two stars it is more likely that their binarity and neutron capture enhancement are unrelated phenomena.

In contrast to the Ba and Y abundances, the [Nd/Fe] values of the Eu-enhanced stars are almost all elevated in comparison to the typical disk sample, including for the r-II stars. This result is especially interesting given that Nd has a much higher $r$-process contribution at Solar metallicities than either Ba or Y, suggesting that these stars are enhanced not only in Eu but may in fact be $r$-process enhanced across multiple abundances. Our findings with regard to Nd are also consistent with the work of \citet{Xie_2025}, who find their two r-II disk stars to also be Nd-rich compared to other metal-rich stars in the Galaxy (see Fig.~5 of that work). The two Ba-enhanced stars also remain the most Nd-enhanced, perhaps because they have had both a high $s$- and $r$-process contribution to increase the [Nd/Fe] abundance or because they are especially $s$- or $i$-process rich, which produces slightly more Nd than the $r$-process.

The right panel of Fig.~\ref{fig:neutron_capture} shows the [Y/Ba] abundances of our target stars relative to the overall disk distribution, thus giving an approximation of light-to-heavy $s$-process element abundances. This ratio will trace AGB masses of $s$-process enrichment sources, with low-mass AGBs producing more heavy $s$-process elements (e.g., Ba) and high-mass AGBs producing lighter elements \citep[e.g., Y;][]{Karakas_Lattanzio_2014,Karakas_Lugaro_2016}. As is evident in the figure, almost all of the Eu-rich target stars have [Y/Ba] abundances that appear approximately typical for disk stars. This includes the at least one of the two $s$-process enhanced target stars, which has $\textrm{[Y/Ba]}=-0.24$; the other has a slightly lower $\textrm{[Y/Ba]}=-0.42$. If these two stars were enriched by an AGB companion, this could be an indication of differing masses for that companion between the two.

\subsection{Comparison to Other Systems}
\label{subsec:comparison}

These stars were identified for r-II levels of Eu enhancement, a criterion typically used in studies of metal-poor stars. Although this selection is clearly less empirically motivated for metal-rich disk stars, as we discuss further in Section~\ref{subsec:disk_enhanced_selection}, we have used it here to enable comparisons of our sample to the multitude of previous studies of $r$-process enhanced stars, which most often focus on metal-poor stars. Thus, in this section, we make those comparisons which initially motivated the selection of our sample for this work. In particular, we are looking for the ways in which our Eu-enhanced disk stars both compare to and differ from their counterparts at lower metallicities, across a selection of elements beyond Eu alone.

In Fig~\ref{fig:abundance_comparison}, we compare the Mg, Eu, and Ba abundances of our target stars to MW halo stars from the $r$-Process Alliance \citep[RPA][]{Ezzeddine_2020} and members of classical dwarf spheroidal galaxies (dSphs) from a selection of literature sources \citep{norris2017b, hill2019, theler2020, reichert2020}. Although the the RPA stars are generally metal-poor, consistent with the fact that most $r$-process enhanced stars are themselves low metallicity, the dSphs populate metallicities up to almost Solar, providing valuable \textit{ex-situ} benchmarks against which to compare the Eu-enhanced disk stars. In the left panel, the [Mg/Fe] abundances of our target stars in this work are clearly elevated compared to those of the dSphs at similar metallicities, reflecting the higher star formation rates of the MW compared to these much smaller systems. By contrast, the dSph members appear to experience the $\alpha$-element ``knee'' at much lower metallicities, indicating the dominance of Fe production from Type Ia SNe rather than CCSNe in periods of high star formation. The clearly distinct Mg abundances of the target stars suggest strongly that in spite of their unusual neutron capture abundances, these are true \textit{in situ} disk stars.

The second panel of Fig~\ref{fig:abundance_comparison} compares the [Ba/Fe] abundances of our target stars to those from the RPA stars and the dSph members. The target stars are not so distinct from the metal-rich dSph members in Ba as they were in Mg, occupying largely same space as the upper end of the [Ba/Fe] distribution of the dwarf galaxy stars near Solar metallicity. The two very Ba-enhanced, Eu-rich disk stars identified in Fig.~\ref{fig:neutron_capture} are again evident here; notably, several of the RPA halo stars appear to have similar levels of enrichment in [Ba/Fe] to these two disk stars. These Ba-enhanced RPA stars are also very Eu-rich, all having $\mathrm{[Eu/Fe]}\geq0.9$. Correspondingly, the last panel of the figure, which shows [Eu/Ba] versus [Fe/H], demonstrates that several of the RPA stars also have low [Eu/Ba] abundances, like the two unusual disk stars and distinct from the bulk of the RPA [Eu/Ba] abundances. The parallels in their neutron capture abundances could suggest analogous enrichment sources or nucleosynthetic pathways across a range of metallicites, although this is purely speculative. 
The third panel of Fig.~\ref{fig:abundance_comparison} shows the [Eu/Fe] versus [Fe/H] abundances of the Eu-rich disk stars and dwarf galaxy member stars. As a result of the $\mathrm{[Eu/Fe]}>0.7$ selection line which is imposed upon all of our target stars, the target stars are here again notably enhanced in Eu relative to the dGal members. The bulk of the metal-rich dSph stars sit below the r-II selection, and correspondingly, most of r-II stars among these dwarf galaxies fall below $\mathrm{[Fe/H]}<-1$, much like in the MW. Interestingly, however, the classical dSph galaxies do host three Eu-enhanced stars with $\mathrm{[Eu/Fe]}>0.7$ at $\mathrm{[Fe/H]}>-1$, suggesting that this degree of $r$-process enrichment is not unique to the Milky Way disk at high metallicities. Interesting future work might explore the occurence rates of these stars across enviroments as a way to constrain rates of neutron capture events.

\section{Discussion}
\label{sec:discussion}

\subsection{The Site of Europium Production in the Galactic Disk}
\label{subsec:enrichment_source}

\begin{figure}
    \centering\includegraphics[width=\columnwidth, alt={The Eu-enhanced stars have much higher [Eu/Mg] than the rest of the disk stars. In the left panel, the disk stars form a cloud centered around [Eu/Mg]=0, with most between about -0.4 and 0.3, and [Eu/Mg] decreases slightly with increasing metallicity, from about 0.1 at [Fe/H]=-1 to about -0.2 at [Fe/H]=0.3. By contrast, all but two of the Eu-enhanced stars have [Eu/Mg] greater than 0.5, with values between about 0.42 and 0.95, and they sit clearly above the disk distribution at [Fe/H] between about -0.9 and -0.2. The most Eu-enhanced of the plotted stars, at [Fe/H] of about -0.5, also has the highest [Eu/Mg], of about 0.95. In the right panel, which shows [Eu/Mg] versus [Mg/Fe], the disk stars are concentrated near [Mg/Fe]=0, and the Eu-enhanced stars have [Mg/Fe] of about -0.05 to 0.28, similar to typical disk stars. There is no clear correlation between [Eu/Mg] and [Mg/Fe] among the Eu-enhanced stars, meaning they are enhanced in Eu without a corresponding enhancement in Mg. The typical uncertainty is about 0.13 in [Eu/Mg], and less than 0.1 [Fe/H] and [Mg/Fe].}]{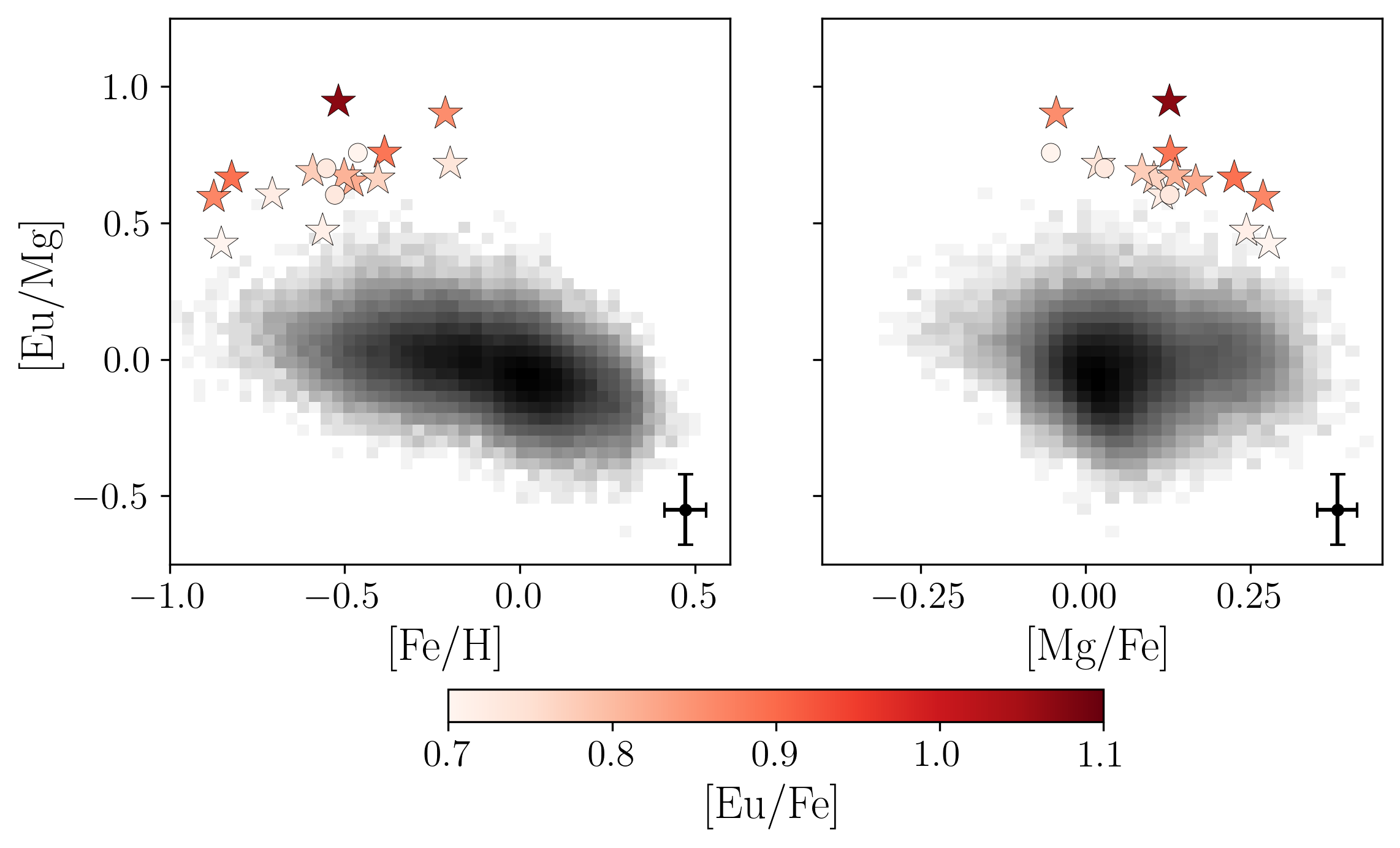}
    \caption{The distribution of [Eu/Mg] versus [Fe/H] (left) and [Mg/Fe] (right). The gray background histogram shows the distribution of all selected disk stars in GALAH DR4. Overplotted with star or circle markers are the selected r-II or other Eu-enhanced disk stars, color-coded by their [Eu/Fe] abundances. Note the high level of [Eu/Mg] elevation among the overall Eu-enhanced sample.}
    \label{fig:eu_mg}
\end{figure}

\begin{figure*}
    \centering\includegraphics[width=1.55\columnwidth, alt={The Eu-enhanced stars have much higher [Eu/Mg] than the rest of the disk stars. In the left panel, the disk stars form a cloud centered around [Eu/Mg]=0, with most between about -0.4 and 0.3, and [Eu/Mg] decreases slightly with increasing metallicity, from about 0.1 at [Fe/H]=-1 to about -0.2 at [Fe/H]=0.3. By contrast, all but two of the Eu-enhanced stars have [Eu/Mg] greater than 0.5, with values between about 0.42 and 0.95, and they sit clearly above the disk distribution at [Fe/H] between about -0.9 and -0.2. The most Eu-enhanced of the plotted stars, at [Fe/H] of about -0.5, also has the highest [Eu/Mg], of about 0.95. In the right panel, which shows [Eu/Mg] versus [Mg/Fe], the disk stars are concentrated near [Mg/Fe]=0, and the Eu-enhanced stars have [Mg/Fe] of about -0.05 to 0.28, similar to typical disk stars. There is no clear correlation between [Eu/Mg] and [Mg/Fe] among the Eu-enhanced stars, meaning they are enhanced in Eu without a corresponding enhancement in Mg. The typical uncertainty is about 0.13 in [Eu/Mg], and less than 0.1 [Fe/H] and [Mg/Fe].}]{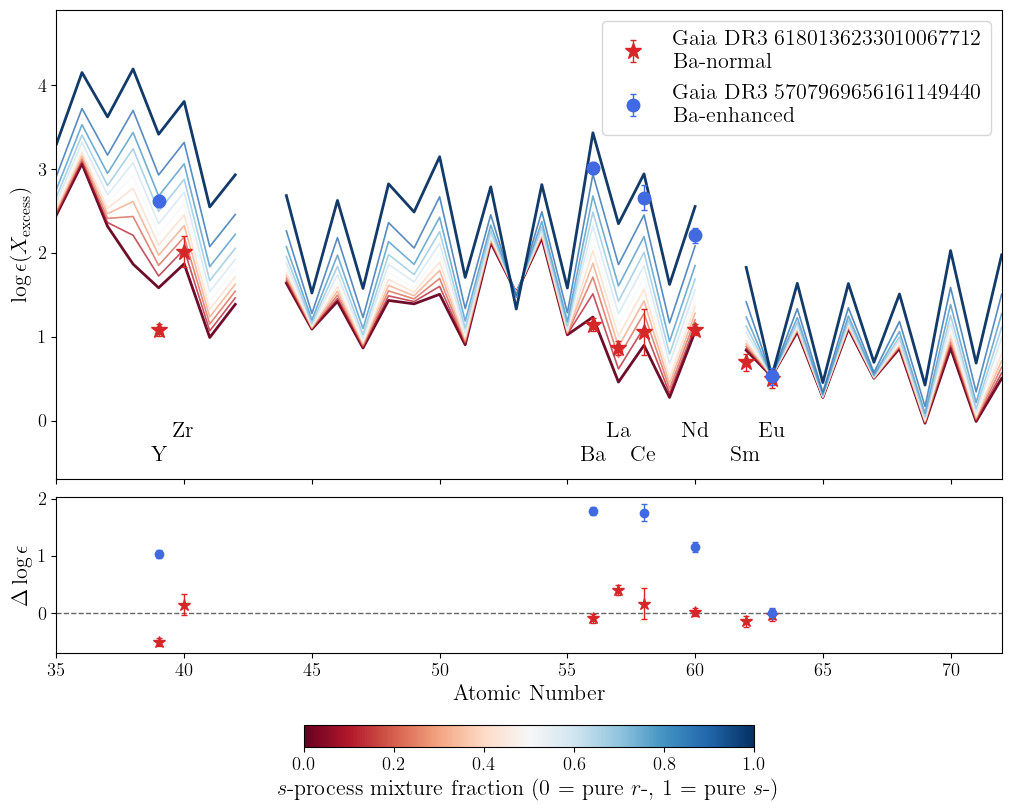}
    \caption{In the top panel are the Solar ratios of the Solar $s$-to-$r$-process patterns, ranging from 0 (pure $r$-process) in red to 1 (pure $s$-process) in blue. $s$- and $r$-process fractions for each element are adopted from \citet{Prantzos_2020} and are applied to the Solar abundances from \citet{Grevesse_2007}. Values are given as log-scaled absolute abundances, and elements of particular interest are labeled. As blue circles, we show the log-scaled absolute abundances of the Eu- and Ba-enhanced star \textit{Gaia} DR3 \texttt{5707969656161149440}, which has $\mathrm{[Fe/H}]=-0.55$ and $\mathrm{[Eu/Fe}]=0.73$. The red star markers indicate the log-scaled absolute abundances of a r-II star from our sample with a similar Eu abundance to \textit{Gaia} DR3 \texttt{5707969656161149440}, with $\mathrm{[Fe/H}]=-0.56$ and $\mathrm{[Eu/Fe}]=0.72$. For each element, we have subtracted the median abundance among disk stars in GALAH DR4 within a $0.1$~dex window of [Fe/H] from the target star, thus making the abundances shown an ``excess'' above the typical values in the disk rather than the true abundance given by GALAH. Only elements in the GALAH DR4 catalog without quality flags raised are used. The Solar $s$-/$r$-process mixtures from \citet{Prantzos_2020} are scaled to the Eu abundance of \textit{Gaia} DR3 \texttt{5707969656161149440}. In the bottom panel, we show the per-element deviation from the pure Solar $r$-process abundances for each star.}
    \label{fig:solar_rprocess}
\end{figure*}

As we have attempted to establish in the previous section, our 17 Eu-enhanced stars appear from their chemistry to be genuine \textit{in-situ} disk stars, which is also corroborated by their orbital properties (see Fig.~\ref{fig:selection}). Thus, identifying the source of their unique Eu enrichment could help shed light on the sites of the neutron capture processes in the MW disk. This is an especially crucial task given that the $r$-process remains arguably the least well-constrained nucleosynthetic channel, with analyses of Eu abundances in the Galactic disk finding scenarios that range from a significant $r$-process contribution from binary NSMs \citep{Schonrich_Weinberg_2019} all the way down to NSMs constituting only a small to moderate fraction of the $r$-process budget \citep{Chen_2025}. Moreover, with the sites of the $r$-process often being cited as metallicity-dependent \citep{Cote_2019, Haynes_Kobayashi_2019, Fraser_2022}, metal-rich tracers of this nucleosynthetic channel become essential alongside their metal-poor counterparts, which are now numerous from other efforts.
The need for further characterization of $r$-process sites is compounded by the dearth of gravitational wave observations of binary NSMs, of which only two have been discovered.\footnote{Although many more were anticipated prior to LIGO beginning its observing runs, underscoring the need for further constraints on NSM rates.} Answering the question of where the $r$-process occurs, and how frequently--in binary NSMs, in certain CCSNe, or elsewhere--is thus of broad astrophysical importance.

In addition to the fact that most $r$-process enhanced stars are found at low metallicities, another essential motivation for examining the $r$-process in the metal poor regime is that many of these stars can be assumed to be $r$-process enriched by a single, or at least relatively few, sources. With the younger, metal-rich stars that constitute our sample, this is not a reasonable assumption for us to make; in our stars' histories is probably a patchwork of Eu enrichment over many generations of star formation. This hypothesis is supported by the [Ba/Eu] abundances for these r-II stars, of which $\sim50\%$ are $>-0.53$, suggesting an incomplete $r$-process dominance compared to the Solar $r$-process, which has $\mathrm{[Ba/Eu]}\approx0.9$ \citep{Prantzos_2020}. Moreover, because the stars span a range of energies within the disk (see Fig~\ref{fig:selection}), have different metallicities (Fig.~\ref{fig:feh_dist}) and other chemical abundances (Figs.~\ref{fig:alpha_elements}~and~\ref{fig:neutron_capture}), it seems probable that they are the descendants of different Eu-production events. These events may have been entirely different sources (e.g., binary NSMs for some stars but not others), further complicating the picture.
Thus, this section is intended solely as a speculative first examination of the possible origins of the Eu-enhancement in these stars.

\subsubsection{CCSNe and $\alpha$-element co-production with the $r$-process}

We begin by examining the $\alpha$-element Mg, produced largely in core-collapse supernovae \citep[CCSNe; including magneto-rotational SNe, ][]{Reichert_2021, Reichert_2023} but not in binary NSMs. In Fig.~\ref{fig:eu_mg}, we show the [Eu/Mg] abundances of the Eu-enhanced targets and of the overall sample of disk stars in GALAH relative to both [Fe/H] and [Mg/Fe]. Here, the [Eu/Mg] abundances of the target stars are clearly much higher than those of typical stars in the MW disk, and all but two of the target stars have $\mathrm{[Eu/Mg]}>0.5$. The elevated [Eu/Mg] abundances of the r-II stars suggest that they are strongly $r$-process enhanced without any substantial corresponding $\alpha$-element enhancement relative to the average disk star, which is also evident in Fig.~\ref{fig:alpha_elements}, where the $\alpha$-element abundances of the r-II stars seem generally consistent with those of either the MW thin or thick disk. In complement to this observation, there is also no clear correlation between increasing Eu and Mg abundances.

To determine whether these Mg abundances can be used as a diagnostic for the presence of CCSNe enrichment, we approximate the level of Mg enhancement that could reasonably be expected from a single event. Accordingly, we follow \citet{Magg_2020} to make a rough calculation of the abundance of a star which forms from gas enriched by a CCSNe after homogeneously mixing with a dilution mas $M_\mathrm{dil}$ in the ISM. 
Adapting from Eq.~5 of \citet{Magg_2020}, the final Solar-scaled abundance [X/H] for an element X can be defined as:
\begin{equation}
    \mathrm{[X/H]}=\log_{10}\frac{M_\mathrm{X,~SNe}+M_\mathrm{X,~ini}}{\mu_X X_H M_\mathrm{dil}} - \log_{10}\frac{N_X,\odot}{N_H,\odot}
    \label{eq:dilution}
\end{equation}
where $\mu_X$ is the atomic mass of element X and $X_H$ is the fraction of H in the gas. $M_\mathrm{X,~SNe}$ is the mass of element X yielded by the SNe, and $M_\mathrm{X,~ini}$ is the initial mass of element X present in the dilution mass $M_\mathrm{dil}$ prior to the SNe, which we calculate from an assumed initial abundance in each element X.
$\log_{10}\frac{N_X,\odot}{N_H,\odot}$ is the Solar abundance of element X, which we take from \citet{Grevesse_2007}.
We assume a dilution gas mass of $10^5~\textrm{M}_\odot$ in accordance with the minimum dilution mass suggested in \citet{Reichert_2023}, and we also use the SNe yields from \citet{Reichert_2023} which include $r$-process production. The metallicity of the progenitor star in  \citet{Reichert_2023} is $Z=0.1Z_\odot$, so for the gas into which the yields are mixed, we adopt a metallicity of $\mathrm{[Fe/H]}=-1$ as well as initial abundances of $\mathrm{[Mg/Fe]}=0.25$ and $\mathrm{[Eu/Fe]}=0.4$, to approximately match observed patterns in these elements. All other initial abundances in the dilution gas are set to Solar at $\mathrm{[X/Fe]}=0$ for each element X. After homogeneously mixing the SNe yields from \citet{Reichert_2023} into $10^5~\mathrm{M}_\odot$ of this gas, the change in the final Mg abundance is $\Delta\mathrm{[Mg/Fe]}\approx0.01$, which is below the level of the abundance errors which could typically be expected from spectra of this resolution and SNR. Even if these stars were enriched in Mg from a CCSNe alongside the Eu enhancement, we would not be able to detect that enrichment here, nor would it be distinguishable from the intrinsic Mg spread among disk stars.
Thus, we show this figure and note the lack of correlation between Eu and Mg abundances in our target stars for completeness but do not interpret it further as a diagnostic to differentiate between CCSNe and other Eu sources.

\subsubsection{The $s$- and $r$-processes as Eu production mechanisms}

Throughout this work thus far, we have often referred to $r$-process enrichment in particular, but as is discussed in relation to Fig.~\ref{fig:neutron_capture}, there does appear to be some non-negligible contribution from the $s$- or $i$-processes to the abundance patterns of at least some of the Eu-enhanced stars. Studies of metal-poor $r$-process enhanced stars have suggested the existence of a ``universality'' among the ratios between $r$-process element abundances \citep{Westin_2000, Frebel_2018, Cowan_2021, Roederer_2022, Racca_2025, Agnos_2026}, often examined with regard to Solar $s$- and $r$-process abundances \citep[][among others]{Arlandini_1999, Sneden_2008, Bisterzo_2014, Prantzos_2020}, which we leverage here to further interrogate the sources of the Eu enhancement among our sample. However, as is mentioned above, our stars are unlikely to exhibit pure $r$-process patterns due to their extended nucleosynthetic enrichment histories. We take this into account in two ways: First, rather than comparing to only a pure $r$-process pattern, we compare to a range of $s$- to $r$-process ratios, ranging from pure $r$-process to pure $s$-process. These ratios are taken following from \citet{Prantzos_2020}, where the corresponding $s$- and $r$-process fractions of each element are applied to the Solar abundances from \citet{Grevesse_2007}. Second, when comparing stars from our Eu-enhanced sample to these abundance patterns, we linearly subtract off the median abundance of each element among disk stars in GALAH DR4 within a 0.1~dex window of [Fe/H] around the target star. In this way, we are examining a ``neutron capture excess'' for our stars to determine whether an $s$-process pattern, $r$-process pattern, or a mixture of the two best explains the deviation from typical disk star abundances.

Fig~\ref{fig:solar_rprocess} illustrates the disk-subtracted neutron capture abundance patterns of two of our target stars relative to the Solar patterns from \citet{Prantzos_2020}, all of which are scaled to the observed Eu abundances of our stars. For comparative purposes, the figure shows one of the Ba-rich, Eu-enhanced stars and also a Ba-typical, r-II star with a similar absolute Eu abundance. These two stars are \textit{Gaia} DR3 \texttt{5707969656161149440} and \texttt{6180136233010067712}, respectively. \textit{Gaia} \texttt{6180136233010067712}, the ``typical'' r-II disk star (with its disk-subtracted abundances marked via red stars), is generally more consistent with mostly $r$-process enrichment, with its Ba, La, Ce, Nd, and Sm abundances all approximately compatible with a $\sim80-100\%$ $r$-process ratio. There is some variation among these elements regarding which $r$-/$s$-process pattern is most favored, which could perhaps be resolved with higher precision abundances, but overall most of the excess neutron capture abundances of \textit{Gaia} \texttt{6180136233010067712} are consistent with a strong $r$-process source.
However, unlike the other neutron capture abundances derived by GALAH for this star, Y is underabundant relative to even the pure the Solar $r$-process by $\sim0.5$~dex. We note that this discrepancy was exacerbated by the subtraction of the disk median abundances, although Y was underabundant even prior to this subtraction. Such underabundance in Y compared to models is an observed issue in multiple works on the $r$-process \citep[e.g., see the discussion in][which suggests that the discrepancy may be reduced by comparing to the \citealp{Bisterzo_2014} abundance patterns]{Ji_Drout_Hansen_2019}, although we do not discount the possibility that our analysis also contributes to the disagreement.

In contrast to the r-II star, the Ba-enhanced star \textit{Gaia} \texttt{5707969656161149440} (shown as blue circle markers in Fig.~\ref{fig:solar_rprocess}) deviates significantly from the Solar $r$-process abundance pattern, with Y, Ba, Ce, and Nd all overabundant by a substantial $1-2$~dex from the $100\%$ $r$-process model. Instead, this star is much more consistent with a $90-100\%$ $s$-process contribution, which is coherently observed across multiple neutron capture elements. Of the heavy elements determined in this star by GALAH DR4, only Y differs substantially from this mostly $s$-process pattern; much like the r-II star, \textit{Gaia} \texttt{5707969656161149440}'s Y abundance is $\sim0.5$~dex below the values suggested by the patterns consistent with the other neutron capture elements. Again, this may be due to measurement errors, or to incompatibility with the \citet{Prantzos_2020} models.
However, based on the abundances of its other neutron capture elements, we suggest that \textit{Gaia} \texttt{5707969656161149440} could plausibly be Eu enhanced via the $s$-process, most likely from its binary companion. Importantly, however, we do not compare the Ba- and Eu-enhanced stars' abundances to an $i$-process model here \citep[e.g., as is done in][]{Karinkuzhi_2023}, but we consider this a particularly exciting avenue for future work.

In their study of Ba-enhanced stars ($\mathrm{[Ba/Fe]}\geq1$), \citet{Levine_2026} report an occurrence rate of $0.62\%$ of such stars within the GALAH survey, although we note that their sample spans a range of Galactic environments. Regardless, within our sample of Eu-enhanced stars, the $2/17$ ($11.8\%$) of targets with $\mathrm{[Ba/Fe]}\geq1$ represents a much higher occurrence rate. This finding may be indicative of the importance of AGB binary mass transfer for producing very Eu-rich stars at high metallicities, if the AGB companion is indeed responsible for performing the Eu enrichment.
Determining whether the two Ba-enhanced binary stars \textit{Gaia} DR3 IDs \texttt{5707969656161149440} and
\texttt{5251879932203856384} were separately $r$- and $s$-process enriched by different events or whether their Eu enhancement also comes from enrichment via their AGB companion via future work will be crucial for answering this question.

\subsection{New Criteria for Selection of \textit{r}-process Enhanced Stars in the Disk}
\label{subsec:disk_enhanced_selection}

\begin{figure}
    \centering\includegraphics[width=\columnwidth, alt={The 99th percentile [Eu/Fe] values of the disk stars decrease steadily with increasing metallicity and are well described by our linear selection criterion. The disk stars form a dense cloud from the upper left (higher [Eu/Fe], higher [Fe/H]) to the lower right (lower [Eu/Fe], lower [Fe/H]), as in Figure 1. The 99th percentile values, calculated in 0.1 dex bins of [Fe/H], are about 0.66 at [Fe/H]=-0.95, 0.6 at -0.65, 0.45 at -0.35, 0.31 at -0.15, 0.19 at 0.05, and 0.1 at 0.35. The linear selection, [Eu/Fe] greater than -0.465 times [Fe/H] plus 0.235, runs from [Eu/Fe]=0.7 at [Fe/H]=-1 to about 0 at [Fe/H]=0.5 and passes within about 0.1 dex of all of the 99th percentile values; the points between [Fe/H] of about -0.8 and -0.3 fall slightly above the line. For comparison, the r-II threshold at [Eu/Fe]=0.7 lies above the entire distribution of disk stars. The r-I threshold at [Eu/Fe]=0.3 matches the 99th percentile only near [Fe/H]=-0.15: it would select a large fraction of the most metal-poor disk stars but very few stars at super-Solar metallicities. Our selection identifies 915 stars, or 1.22 percent of the disk sample (see Section 4.2).}]{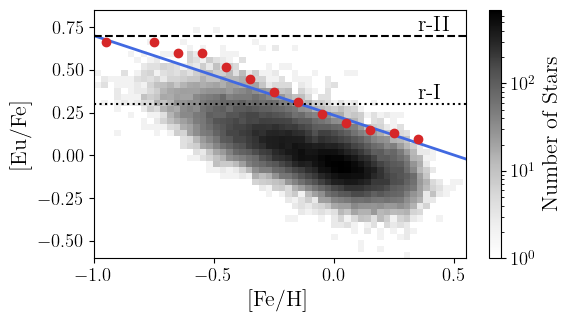}
    \caption{As in the left panel of Fig.~\ref{fig:selection}, the grey histogram shows the log-scaled [Eu/Fe] versus [Fe/H] distribution of disk stars in GALAH DR4. The red points indicate the $99$th percentile [Eu/Fe] abundance at each metallicity, and the blue solid line marks the linear selection we suggest for Eu-enhanced stars. For reference, the r-I and r-II selection criteria at $\mathrm{[Eu/Fe]}=0.3$ and $0.7$ are marked as black dotted and dashed lines, respectively.}
    \label{fig:new_selection}
\end{figure}

As is evident in Figs.~\ref{fig:selection} and \ref{fig:feh_dist} and has been frequently noted in previous works \citep{Battistini_Bensby_2016,Zhao2016}, the [Eu/Fe] abundance of stars with $\mathrm{[Fe/H]}>-1$ anticorrelates strongly with metallicity. Thus, the r-II selection criterion for $\mathrm{[Eu/Fe]}>0.7$, which is empirically well-motivated in the halo where [Eu/Fe] abundances are approximately constant with respect to metallicity and generally higher than in the disk, is less well-suited in the metal-rich regime. Thus, although we have chosen this r-II selection to benchmark this work against halo studies of $r$-process enhanced stars, future works should consider better-suited selection criteria in the disk. The identification of these enhanced stars may be useful to characterize the sites of the $r$-process in the metal-rich regime.


We propose here a first attempt at selecting Eu-rich stars in the disk, the goal of which is to identify the stars with the highest [Eu/Fe] as a function of metallicity. As such, we begin by splitting our sample into $0.1$~dex bins of [Fe/H] from $-1$ to $+0.4$ and calculating the $99$th percentile of the [Eu/Fe] abundances in each of these bins. This $99$th percentile mark could itself be an empirical selection for Eu-enhanced, metal-rich stars, but for more convenient use, we represent these $99$th percentile values as a simple best-fit linear model, weighted by the number of stars in each metallicity bin. We also force the model to reach $\mathrm{[Eu/Fe]}=0.7$ at $\mathrm{[Fe/H]}=-1$, thus ensuring convergence with the r-II selection in the metal-poor regime. Usefully, the $99$th percentile [Eu/Fe] abundances at $\mathrm{[Fe/H]}=-1$ fall naturally close to $0.7$. The simple linear selection for Eu-enhanced stars is:
\begin{equation}
    \mathrm{[Eu/Fe]}>-0.465\times\mathrm{[Fe/H]}+0.235
    \label{eq:selection}
\end{equation}

This selection, illustrated in Fig.~\ref{fig:new_selection}, identifies $915$ stars ($1.22\%$ of the disk sample) from the overall sample of disk stars in GALAH DR4. This is a deliberately lower rate (i.e., because we select only the top $1\%$ of stars) of Eu-enhanced stars than the r-II occurrence rate in the metal-poor halo, which is closer to $12\%$ \citep{Hansen_2018, RPA_Sakari_2018, Ezzeddine_2020}, reflecting the lower spread in [Eu/Fe] abundances in the disk \citep{AMBRE_Guiglion_2018, DelgadoMena_2017}. Other surveys such as the 4-metre Multi-Object Spectroscopic Telescope \citep[4MOST,][]{4MOST_deJong_2019,4MOST_Walcher_2019} which include many Eu abundances for metal-rich stars could re-derive these selection criteria for the top $1\%$ of [Eu/Fe] values in each metallicity bin of $0.1$~dex, which we suggest to be worthwhile as the spread in the distribution could reasonably be driven by measurement uncertainties. Similarly, we suggest that these outlier [Eu/Fe] abundances be confirmed individually, as in large-scale surveys, many such abundances from the pipeline may be spurious, as is discussed in Section~\ref{subsec:synthesis}.

We also propose that true $r$-process enhanced disk stars ought to have $\textrm{[Ba/Eu]}<-0.3$ to ensure stronger $r$- than $s$-process enrichment. This requirement corresponds to a ratio of Ba to Eu approximately half that of the Sun, which is more strict than the $\textrm{[Ba/Eu]}<0$ requirement for r-I and r-II stars. We suggest that this more stringent requirement may be prudent for metal-rich stars given that they likely will have experienced a long enrichment history that includes both $r$- than $s$-process sources. Of the 14 r-II disk stars identified in this work, 12 have $\textrm{[Ba/Eu]}<-0.3$, and of the $915$ stars identified with the metallicity-dependent [Eu/Fe] selection above, $615$ additionally have $\textrm{[Ba/Eu]}<-0.3$. 
Alternatively, the criterion involving Ba may be omitted altogether if one is interested in Eu-enhanced stars which also appear $s$- or $i$-process enriched, as we see for several binary stars in our sample.


\section{Conclusions}
\label{sec:conclusions}

In this study, we presented a selection of $17$ newly discovered Eu-enhanced metal-rich, disk stars with spectroscopically confirmed $\mathrm{[Eu/Fe]}>0.7$ at $\mathrm{[Fe/H]}>-1$. Fourteen of these stars have $\mathrm{[Ba/Eu]}<0$, making them r-II stars. This sample represents the largest homogeneously analyzed collection of Eu-enhanced, r-II candidate stars at high metallicities, which is made possible by the scale and good Galactic disk coverage available in GALAH DR4 \citep{GALAH_DR4}. We summarize our findings as follows:
\begin{itemize}
    \item The r-II candidates span both the high and low-$\alpha$ disks and otherwise appear approximately typical in $\alpha$-element abundances compared to other disk stars in GALAH DR4 (see. Fig~\ref{fig:alpha_elements}). They are chemically distinct from members of classical dwarf spheroidal galaxies (Fig.~\ref{fig:abundance_comparison}), which, taken in combination with our kinematic selection criteria, strongly suggests that most or all of the target stars here are genuine MW disk members.
    \item Most of our Eu-enhanced candidates are approximately typical in the $s$-process tracer elements Ba and Y compared to other disk stars. Their $r-$ to $s$-process abundances are thus elevated, with a median [Ba/Eu] value of $-0.53$. By contrast, essentially all of our target stars are enhanced in Nd, which has a higher $r$-process contribution at Solar metallicity than either Ba or Y (see Fig.~\ref{fig:neutron_capture}).
    \item Two target stars are very enhanced in Ba and Y in addition to Eu and Nd. The astrometry and radial velocities of these targets suggest that the source of their $s$-process enhancement is binary mass transfer from an AGB companion. These two stars are prime targets for future spectroscopic follow-up and characterization with more heavy elements.
    \item The target stars lack any obvious correlation between their Mg ($\alpha$-element) and Eu abundances (Fig.~\ref{fig:eu_mg}). The stars thus appear to be enhanced solely in the neutron capture elements and, in particular, $r$-process tracer elements for most stars. However, the long and complex enrichment histories in the disk mean that identifying which source produced the Eu enrichment--particularly, distinguishing between NSMs and CCSNe--is not possible at this stage.
    \item Comparison of the abundances of two of our target stars with the Solar $s$- and $r$-process models from \citet[][Fig.~\ref{fig:solar_rprocess}]{Prantzos_2020} suggest that the r-II stars may be enriched beyond typical disk neutron capture abundances by a predominantly $r$-process source, whereas the Ba-enhanced, Eu-enhanced binary star shows more consistency with $s$-process enrichment.
    \item Although we select stars with the r-II criterion in Eu used for the halo to situate our work within previous literature studies, a metallicity-independent cut for $r$-process is not as well-suited at high metallicities, where [Eu/Fe] clearly evolves with [Fe/H] (see the left panel of Fig.~\ref{fig:neutron_capture}). Thus, we propose a new selection criterion for $r$-process enhanced disk stars, which is designed to select the upper $1\%$ [Eu/Fe] values as a function of metallicity. This selection is $\mathrm{[Eu/Fe]}>-0.465\times\mathrm{[Fe/H]}+0.235$, although it could be modified slightly for other surveys. We also suggest a $\mathrm{[Ba/Eu]}<-0.3$ requirement, for a ratio of Ba to Eu half that of the Sun. We suggest that these new selection criteria may be especially interesting in light of upcoming stellar spectroscopic surveys like 4MOST \citep{4MOST_deJong_2019,4MOST_Walcher_2019}.
\end{itemize}

We are following up several of the Eu-enhanced disk stars with higher resolution, higher signal-to-noise spectroscopy. Further observations will allow both the confirmation of the Eu abundances via multiple lines as well as other heavy \textit{r}- and \textit{s}-process elements, thus enabling a more robust chemical characterization of the observed stars and, hopefully, more certain identification of the progenitors of their \textit{r}-process enrichment. This work thus serves as a broad-view, population-level analysis of neutron capture enhanced stars in the disk as a precursor to more detailed  spectroscopic studies.

The scale and setup of GALAH DR4 offers an unprecedented opportunity to study the $r$-process in the Galactic disk. With the sites of the $r$-process remaining the least well-constrained of all nucleosynthetic channels, this opportunity is invaluable. Here, we have only scratched the surface of the science possible in this regime by providing a preliminary analysis of $r$-process enhanced stars in the disk as a companion to numerous such studies existing in the literature about metal-poor halo stars. Many exciting further avenues of work lie ahead, both in the further characterization of Eu-enhanced stars in the disk and in the bulk distributions of Eu among metal-rich stars. In particular, refinement of the selection of Eu-enhanced stars in the disk as described in Section~\ref{subsec:disk_enhanced_selection} has the potential to enable the use of enhanced stars to constrain $r$-process rates (and thus, potentially NSM merger rates) and yields in the disk. We look forward to such future work.

\section*{Acknowledgements}

SGK thanks Chiaki Kobayashi and Alexander Ji for useful discussions regarding Eu enrichment sources and Jaden Levine for discussions regarding Ba enhancement. 
SGK thanks the Marshall Scholarship for PhD funding, with joint support from the UK government, the Cambridge Trust, and Trinity College, Cambridge. Support for this work was provided by NASA through the NASA Hubble Fellowship grant \#HST-HF2-51560 awarded by the Space Telescope Science Institute, which is operated by the Association of Universities for Research in Astronomy, Inc., for NASA, under contract NAS5-26555. T.T.H. acknowledges support from the Swedish Research Council (VR 2025-05377).

This work made use of the Fourth Data Release of the GALAH Survey (Buder et al. 2021). The GALAH Survey is based on data acquired through the Australian Astronomical Observatory, under programs: A/2013B/13 (The GALAH pilot survey); A/2014A/25, A/2015A/19, A2017A/18 (The GALAH survey phase 1); A2018A/18 (Open clusters with HERMES); A2019A/1 (Hierarchical star formation in Ori OB1); A2019A/15, A/2020B/23, R/2022B/5, R/2023A/4, R2023B/5 (The GALAH survey phase 2); A/2015B/19, A/2016A/22, A/2016B/10, A/2017B/16, A/2018B/15 (The HERMES-TESS program); A/2015A/3, A/2015B/1, A/2015B/19, A/2016A/22, A/2016B/12, A/2017A/14, A/2020B/14 (The HERMES K2-follow-up program); R/2022B/02 and A/2023A/09 (Combining asteroseismology and spectroscopy in K2); A/2023A/8 (Resolving the chemical fingerprints of Milky Way mergers); and A/2023B/4 (s-process variations in southern globular clusters). We acknowledge the traditional owners of the land on which the AAT stands, the Gamilaraay people, and pay our respects to elders past and present. This paper includes data that has been provided by AAO Data Central (datacentral.org.au).

This work has made use of data from the European Space Agency (ESA) mission Gaia (https://www.cosmos.esa.int/gaia), processed by the Gaia Data Processing and Analysis Consortium (DPAC, https://www.cosmos.esa.int/web/gaia/dpac/consortium). Funding for the DPAC has been provided by national institutions, in particular the institutions participating in the Gaia Multilateral Agreement.

\section*{Data Availability}

This work is based upon publicly available catalogs from the \href{https://www.galah-survey.org/dr4/overview/}{GALAH survey}. The GALAH DR4 spectra are also publicly available. Astrometry and proper motions from the \textit{Gaia} mission and distances from \citet{BailerJones_2021} are also publicly available. The full list of Eu-enhanced stars discussed in this work are available in Appendix~\ref{appendix}.



\bibliographystyle{mnras}
\bibliography{bibliography} 




\appendix

\section{Sample of Eu-enhanced disk stars}
\label{appendix}

The full list of r-II stars in the Galactic disk used in this work are given in Table~\ref{tab:eu_enhanced_stars}. 

\begin{table*}
\centering
\caption{All $17$ confirmed Eu-enhanced stars in the disk sample. $T_\mathrm{eff}$, $\log g$, [Fe/H], and [Ba/Fe] are given as the GALAH DR4 values, and [Eu/H] abundances are those derived from synthesis with \texttt{Korg}. Those stars which are Eu-enhanced but not r-II (e.g., which have $\mathrm{[Ba/Eu]}>0$) are marked with $^\dagger$.\label{tab:eu_enhanced_stars}}
\scriptsize
\setlength{\tabcolsep}{2pt}
\renewcommand{\arraystretch}{1.08}
\resizebox{\textwidth}{!}{%
\begin{tabular}{rrcccccccccc}
\hline
\shortstack{GALAH\\SObject ID} & \shortstack{Gaia DR3\\Source ID} & RA [$^{\circ}$] & Dec [$^{\circ}$] & \shortstack{$T_{\rm eff}$\\{[K]}} & $\log g$ & $\mathrm{[Fe/H]}$ & $\mathrm{[Eu/H]}$ & $\sigma_\mathrm{[Eu/H]}$ & $\mathrm{[Eu/Fe]}$ & $\sigma_\mathrm{[Eu/Fe]}$ & $\mathrm{[Ba/Eu]}$ \\
\hline
140809002101130 & 5812990677522553088 & 265.512 & -65.542 & 4780 & 2.54 & -0.21 & 0.64 & 0.13 & 0.86 & 0.14 & -0.48 \\
$^\dagger$150211003201276 & 5707969656161149440 & 127.124 & -18.336 & 5433 & 3.79 & -0.55 & 0.18 & 0.1 & 0.73 & 0.11 & 0.69 \\
150831004001105 & 2594341150806513152 & 338.127 & -17.875 & 4753 & 2.64 & -0.52 & 0.55 & 0.12 & 1.07 & 0.13 & -0.63 \\
160126003201308 & 5554733916052133504 & 93.963 & -47.274 & 5000 & 2.86 & -0.87 & -0.01 & 0.1 & 0.86 & 0.12 & -0.64 \\
160425001901055 & 6180136233010067712 & 195.106 & -33.786 & 4774 & 2.57 & -0.56 & 0.15 & 0.1 & 0.72 & 0.12 & -0.59 \\
160813001601332 & 5815617720050615296 & 249.485 & -66.338 & 4893 & 2.48 & -0.39 & 0.5 & 0.11 & 0.89 & 0.12 & -0.53 \\
160816002701040 & 6649156522981897216 & 276.625 & -56.275 & 5261 & 2.77 & -0.85 & -0.15 & 0.09 & 0.7 & 0.12 & -0.76 \\
$^\dagger$170404003601229 & 4170304908161730432 & 264.4 & -5.177 & 4697 & 2.46 & -0.46 & 0.24 & 0.11 & 0.7 & 0.13 & 0.22 \\
$^\dagger$210122003601216 & 5251879932203856384 & 158.028 & -63.784 & 5341 & 3.8 & -0.53 & 0.2 & 0.09 & 0.73 & 0.11 & 0.61 \\
210327004101242 & 5826335450183043712 & 235.735 & -63.433 & 4376 & 1.66 & -0.71 & 0.01 & 0.11 & 0.72 & 0.12 & -0.44 \\
210519003601335 & 4072767025506772736 & 279.154 & -27.258 & 4800 & 2.47 & -0.48 & 0.34 & 0.12 & 0.82 & 0.13 & -0.81 \\
211216002601171 & 3315270000060075264 & 91.544 & 1.762 & 6352 & 3.16 & -0.21 & 1.01 & 0.09 & 1.22 & 0.11 & -0.47 \\
211217002601180 & 3317287569535823616 & 91.723 & 3.836 & 4927 & 3.18 & -0.41 & 0.36 & 0.11 & 0.77 & 0.12 & -0.74 \\
220122001601022 & 3121696446800800768 & 94.215 & -0.099 & 4403 & 1.68 & -0.82 & 0.07 & 0.12 & 0.89 & 0.13 & -0.85 \\
220215003101247 & 5343958808333671296 & 177.792 & -55.436 & 4825 & 2.81 & -0.2 & 0.54 & 0.1 & 0.74 & 0.12 & -0.26 \\
220421002101117 & 5407989761325108992 & 149.13 & -48.615 & 4421 & 1.77 & -0.59 & 0.18 & 0.12 & 0.78 & 0.13 & -0.22 \\
230201001101092 & 5347640518706841344 & 169.117 & -52.992 & 4927 & 2.45 & -0.5 & 0.31 & 0.11 & 0.81 & 0.13 & -0.54 \\
\hline
\end{tabular}
}
\end{table*}

\section{Re-computed Stellar Parameters for the Eu-enhanced Disk Stars}
\label{appendix:stellar_params}

For completeness, we also check the GALAH DR4 $T_\mathrm{eff}$ and $\log g$ values in our sample with photometry.
Photometric effective temperatures are computed following the procedure in \citet{Mucciarelli_2021} using \textit{Gaia} DR3 \texttt{BP-RP} colors and \textit{g} magnitudes \citep{Gaia_DR3_2023}. Metallicities are adopted from GALAH DR4, although as is noted by \citet{Mucciarelli_2021}, uncertainties in [Fe/H] have a minimal effect on the derived temperatures. Extinction values are determined from the dust maps determined in \citet{Schlegel_1998} via interface with the \texttt{dustmaps} package \citet{Green_2018_dustmaps}. The extinction coefficients for \textit{Gaia} photometry were then computed using the \texttt{extinction\_coefficient} package \citep{Zhang_Yuan_2023}. Note that although we have defined all of our stars as giants, those with $\log g>3.0$ are fit with the dwarf photometric $T_\mathrm{eff}$ relation from \citet{Mucciarelli_2021}. $\log g$ values are then re-derived following the relation from Eq.~1 in \citet{Roederer_2018}. We assume masses of $1~\mathrm{M}_\odot$ for all target stars.

To assess the impact of these different $T_\mathrm{eff}$ and $\log g$ values on the derived [Eu/H], we re-compute the abundances with \texttt{Korg} as described in Section~\ref{subsec:synthesis}, now replacing the effective temperature and surface gravity values from GALAH DR4 with those from \citet{Mucciarelli_2021} and \citet{Roederer_2018}, respectively. For most stars, the impact of these differing stellar parameters on the Eu abundance is small; in all but two stars, the difference between the [Eu/H] abundance determined with GALAH DR4 $T_\mathrm{eff}$ and $\log g$ and the re-computed values is $<0.1$.
For easy reference, the re-computed $T_\mathrm{eff}$ and $\log g$ values for the final 17 stars in the sample along with the [Eu/H] abundances determined from synthesis using these new stellar parameters is included in Table~\ref{tab:photometric_sp_eu}.

The photometric temperatures are systematically hotter than those, although for many stars the agreement with the GALAH DR4 reported $T_\mathrm{eff}$ is good. Nonetheless, for six stars, the photometric $T_\mathrm{eff}$ values are $>300$~K greater than those from GALAH DR4. In essentially all of these cases, the extinction values reported by \citet{Schlegel_1998} are high ($\gtrsim0.4$), meaning the photometric temperatures are potentially less reliable. Given that the fits of the GALAH DR4 synthetic spectra appeared to show good agreement with the observations visually, as described in Section~\ref{subsec:synthesis}, we do not believe that the discrepancies between the photometric and GALAH values are entirely real. Nonetheless, we include them here for completeness and to demonstrate that even were the stellar parameters this discrepant, the effect on the derived [Eu/H] abundances is small. $\Delta \mathrm{[Eu/H]}<0.1$~dex for all but two stars in our sample when using the new stellar parameters rather than those from GALAH DR4.

\begin{table*}
\centering
\caption{Provided are the GALAH DR4 and \textit{Gaia} DR3 source IDs as well as the effective temperatures and surface gravities derived following from \citet{Mucciarelli_2021} and \citet{Roederer_2018}, respectively. The [Eu/H] values determined with \texttt{Korg} using these new stellar parameters is also given, alongside the differences in $T_\mathrm{eff}$, $\log g$, and [Eu/H] from the values reported in Table~\ref{tab:eu_enhanced_stars}. $\mathrm{E(B-V)}$ values from \citet{Schlegel_1998} are also provided. \label{tab:photometric_sp_eu}}
\begin{tabular}{|rr|ccccccc|}
\hline
Gaia DR3 Source ID & GALAH SObject ID & $T_{\rm eff,phot}$ [K] & $\Delta T_{\rm eff}$ [K] & $\log g_{\rm phot}$ & $\Delta \log g$ & $\mathrm{[Eu/H]}_{\rm phot}$ & $\Delta \mathrm{[Eu/H]}_{\rm phot}$ & $E(B-V)$ \\
\hline
5812990677522553088 & 140809002101130 & 4784 & 4 & 2.56 & 0.01 & 0.65 & 0.0 & 0.065 \\
5707969656161149440 & 150211003201276 & 5579 & 145 & 3.73 & -0.06 & 0.15 & -0.02 & 0.074 \\
2594341150806513152 & 150831004001105 & 4784 & 31 & 2.71 & 0.08 & 0.58 & 0.03 & 0.046 \\
5554733916052133504 & 160126003201308 & 5056 & 56 & 2.9 & 0.04 & 0.01 & 0.01 & 0.049 \\
6180136233010067712 & 160425001901055 & 4833 & 58 & 2.64 & 0.07 & 0.17 & 0.02 & 0.067 \\
5815617720050615296 & 160813001601332 & 4935 & 42 & 2.53 & 0.05 & 0.52 & 0.02 & 0.09 \\
6649156522981897216 & 160816002701040 & 5282 & 20 & 2.44 & -0.32 & -0.29 & -0.14 & 0.093 \\
4170304908161730432 & 170404003601229 & 4913 & 215 & 2.43 & -0.03 & 0.19 & -0.05 & 0.855 \\
5251879932203856384 & 210122003601216 & 6690 & 1348 & 3.77 & -0.03 & 0.5 & 0.3 & 0.47 \\
5826335450183043712 & 210327004101242 & 4547 & 170 & 1.81 & 0.15 & 0.04 & 0.03 & 0.279 \\
4072767025506772736 & 210519003601335 & 5281 & 480 & 2.43 & -0.04 & 0.28 & -0.07 & 0.393 \\
3315270000060075264 & 211216002601171 & 6845 & 492 & 2.69 & -0.47 & 1.08 & 0.07 & 0.482 \\
3317287569535823616 & 211217002601180 & 5386 & 458 & 3.16 & -0.02 & 0.31 & -0.05 & 0.582 \\
3121696446800800768 & 220122001601022 & 4772 & 368 & 1.83 & 0.14 & 0.07 & 0.0 & 0.381 \\
5343958808333671296 & 220215003101247 & 5169 & 344 & 2.73 & -0.08 & 0.44 & -0.09 & 0.346 \\
5407989761325108992 & 220421002101117 & 4476 & 54 & 1.9 & 0.13 & 0.23 & 0.05 & 0.448 \\
5347640518706841344 & 230201001101092 & 5184 & 257 & 2.46 & 0.01 & 0.29 & -0.02 & 0.245 \\
\hline
\end{tabular}
\end{table*}


\bsp	
\label{lastpage}
\end{document}